\documentclass[12pt,letterpaper]{amsart}

\usepackage{setspace}

\usepackage{amsmath}%
\usepackage{amsfonts}%
\usepackage{amssymb}%
\usepackage[foot]{amsaddr}%

\ifx\pdfoutput\undefined
 \usepackage{graphicx}
\else
 \usepackage[]{graphicx}
 \usepackage{epstopdf}
\fi

\usepackage[numbers]{natbib}
\newlength{\fw}
\begin{document}

\title[Particle trajectory through a constriction]{Analysis of the trajectory of a sphere moving through a geometric constriction}
\author{Sumedh R. Risbud$^{\dagger,*}$}
\address{$^\dagger$Chemical and Biomolecular Engineering, Johns Hopkins University}
%\address{$^*$Mechanical and Aerospace Engineering, Rutgers, the State University of New Jersey}
\author{Mingxiang Luo$^\dagger$}
%\address{Chemical and Biomolecular Engineering, Johns Hopkins University}
\author{Jo\"{e}lle Fr\'{e}chette$^\dagger$}
%\address{Chemical and Biomolecular Engineering, Johns Hopkins University}
\author{German Drazer$^{*,\ddagger}$}
\address{$^*$Mechanical and Aerospace Engineering, Rutgers, the State University of New Jersey}
\address{$^{\ddagger}$ Correspondence: german.drazer@rutgers.edu}

\begin{abstract}
We present a numerical study of the effect that fluid and particle inertia have on the motion of suspended 
spherical particles through a geometric constriction to understand analogous microfluidic settings,
 such as pinched flow fractionation devices. The particles are driven by a constant force in a quiescent 
fluid, and the constriction (the {\em pinching gap}) corresponds to the space between a plane wall and a 
second, fixed sphere of the same size (the {\em obstacle}). The results show that, due to inertia and/or 
the presence of a geometric constriction the particles attain smaller separations to the obstacle. We then 
relate the minimum surface-to-surface separation to the effect that short-range, repulsive non-hydrodynamic 
interactions (such as solid-solid contact due to surface roughness, electrostatic double layer repulsion, 
etc.) would have on the particle trajectories. In particular, using a simple hard-core repulsive potential 
model for such interactions, we infer that the particles would experience larger lateral displacements moving 
through the pinching gap as inertia increases and/or the aperture of the constriction decreases. Thus, 
separation of particles based on differences in density is in principle possible, owing to the differences in 
inertia associated with them. We also discuss the case of significant inertia, in which the presence of a 
small construction may hinder separation by reducing inertia effects.
\end{abstract}

\maketitle
\section{Introduction}\label{sec:intro}
The deterministic motion of spherical particles through geometric constrictions is at the core of a number 
of microfluidic unit operations and devices. Some prominent examples include separation systems, e.g., 
based on pinched flow fractionation,\citep{yamada2004} particle focusing methods,\cite{xuan2010} such as 
geometrical focusing,\citep{faivre2006} and constricted microchannels and micromodels of porous media used 
to study particle deposition and clogging.\cite{wyss2006,mustin2010} In all cases, the interactions of the 
suspended particles with the confining walls of a microfluidic channel play a crucial role in determining 
their motion and fate. The nature of the  interactions can be hydrodynamic, particularly important when 
particle size is comparable to the confinement, or non-hydrodynamic, such as solid-solid contact due to 
surface roughness. 

The motion of suspended particles in microfluidic systems is in many cases described based on the fluid 
flow streamlines obtained in the absence of particles. Rigorously, such an explanation is valid only for 
point particles, because it neglects the effect of particle size and particle-wall hydrodynamic interactions. 
Moreover, the streamlines are calculated in the Stokes regime, neglecting fluid inertia.  Particle inertia 
is also ignored and the effect of non-hydrodynamic interactions, such as solid-solid contact, is typically 
incorporated indirectly. For example, Jain and Posner model the effect of roughness on particle dispersion 
in order to estimate the relevant statistical properties of the separation process, such as its resolution.\citep{jain2008} 
However, there is ample evidence that hydrodynamic interactions, inertia effects, and 
non-hydrodynamic repulsive forces significantly affect the individual particle trajectories 
through narrow constrictions.\citep{mortensen2007,balvin2009,shardt2012} 

Consider the case of pinched flow fractionation (PFF), in which the separative lateral displacement experienced 
by particles of unequal size is achieved by passing the suspension through a constriction, followed by a sudden 
expansion.\citep{yamada2004,takagi2005,sai2006,jain2008,maenaka2008,vig2008} Although the original explanation 
offered for the separative lateral displacement was based on particle motion superimposed on the streamlines in 
Stokes flow,\citep{yamada2004} particle size and hydrodynamic interactions cannot be ignored in PFF due 
to the small particle-wall separations that take place near the constriction.\citep{mortensen2007,shardt2012} 
It has been also shown recently, that inertia effects (manifesting themselves as `lift' forces) might play a 
crucial role in PFF.\citep{shardt2012} In some particle focusing methods inertial effects are also present, and 
even exploited as a suspension of particles moves through a series of constrictions.\citep{park2009,xuan2010} 
In fact, the asymmetric trajectories observed in focusing devices in which the constriction geometry is fore-aft 
symmetric\citep{faivre2006} indicates the presence of {\it irreversible} forces, i.e., inertial forces or 
non-hydrodynamic interactions between the particles and the channel wall. Otherwise, the trajectories should be 
symmetric and reversible, consistent with Stokes equations. In the case of clogging in microchannels, it has 
been argued that particle deposition involves the interplay between hydrodynamic as well as non-hydrodynamic 
interactions.\cite{wyss2006} The hydrodynamic interaction of particles with constrictions/channel walls leads 
to particle trajectories that deviate from flow streamlines, and if a particle reaches a critical separation 
from the wall, non-hydrodynamic attractive interactions result in particle deposition.\cite{wyss2006,mustin2010}

\begin{figure}
\includegraphics[width=1.25\fw]{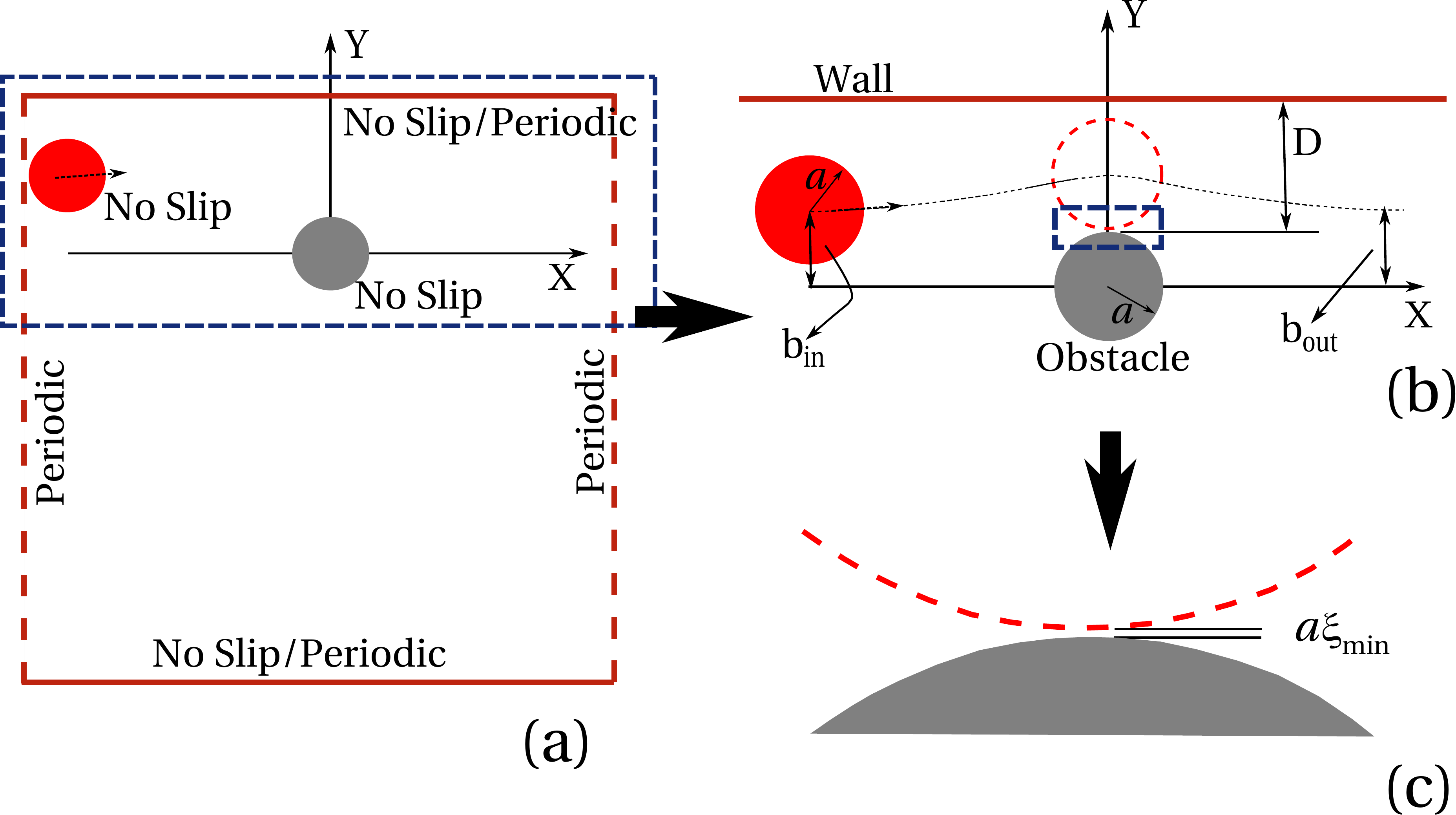}
\caption{(a) System, simulation box and boundary conditions. The schematic is to scale, i.e., the box is 16 
obstacle/particle radii in length. The top wall creates the pinching gap. In the absence of walls, we use the 
periodic boundaries for the box, and the obstacle is at the center of the box. (b) The part of the simulation 
box indicated with dashed lines in (a) is enlarged here. The initial offset $b_{in}$, the final offset $b_{out}$ 
and the aperture of the pinching gap $D$ are indicated. The dashed circle depicts the position of the particle 
when the particle-obstacle separation is minimum (in the absence of inertia). (c) Enlarged view of the particle 
(dashed) and the obstacle (solid) surfaces showing the minimum separation $a \xi_{min}$ between them.}
\label{fig:systemGeometry}
\end{figure}

In this work, we investigate the motion of a spherical particle passing through a constriction created by a 
spherical obstacle of the same size and a plane wall (see figure \ref{fig:systemGeometry}). We are interested in 
the minimum surface-to-surface separation between the particle and the obstacle occuring during the motion, 
since it is the relevant length scale to assess the relative importance of short-range non-hydrodynamic 
interactions. Specifically, we show that the minimum separation decreases as the magnitude of inertia 
increases and/or the aperture of the constriction decreases. The inferences drawn from these results are 
consistent with the observations reported in previous experiments, in that, the lateral displacement of a given 
size of particles increases as the aperture of the constriction becomes smaller.\cite{luo2011} 
In addition, the computational nature of this work enables us to isolate the effects of fluid and 
particle inertia by adjusting both the particle-to-fluid density ratio as well as the force acting on the particle. 
In comparison, the ratio of the densities was moderate in the experiments reported in \cite{luo2011} 
($O(10^0)-O(10^1)$), and hence the effects of particle and fluid inertia were coupled. 

The article is organized as follows: in the following section, we define the geometry of the system under 
investigation and describe the parameter space relevant to the problem in terms of dimensionless numbers. 
In \S\ref{sec:onlyWall}, we introduce a hard-core model for non-hydrodynamic repulsive interactions, and 
discuss its effect on particle trajectories in the absence (\ref{subsec:hardWallModel}) and presence 
(\ref{subsec:hardWallModelInertia}) of inertia. In \S\ref{sec:bothInertiaNdWall}, we present the results 
of our simulations in the presence of a pinching wall only (\ref{subsec:applModelOnlyWall}), in the presence 
of inertia only (\ref{subsec:applModelOnlyInertia}), and in the presence of both a pinching wall as well as 
significant inertia (\ref{subsec:applModelBoth}). Finally in \S\ref{sec:summary} we summarize the inferences 
drawn from this work.

\section{System definition and parameter space}\label{sec:sysDef}
A schematic view of the system under consideration is shown in figure \ref{fig:systemGeometry}. A spherical particle 
suspended in a quiescent fluid moves past a spherical obstacle of the same radius $a$ (diameter $d$) under the action 
of a constant force. The $x$-axis is set in the direction of the constant force, and the far upstream (downstream) -- 
i.e., $x \to -\infty$ ($x \to +\infty$) -- position of the particle along the $y$-axis is the {\it initial offset} 
$b_{in}$ ({\it final offset} $b_{out}$). A plane wall perpendicular to the positive $y$-axis creates a wall-obstacle 
{\it pinching gap} of minimum aperture $D$. The motion of a particle initially located in the $xy$-plane is confined 
to that plane, due to the symmetry of the problem. The minimum value of the dimensionless surface-to-surface separation 
between the particle and the obstacle for a given trajectory is $\xi_{min}$ (the corresponding dimensional separation 
$a\xi_{min}$ is shown in figure \ref{fig:systemGeometry}(c)). 

Numerical simulations are performed using the lattice Boltzmann method ({\em susp3d}).\citep{ladd1994a,ladd1994b,nguyen2002} 
The boundary conditions imposed in the simulations are shown in figure \ref{fig:systemGeometry}(a): no-slip at all fluid-solid 
interfaces, periodic boundary conditions in $x$- and $z$-directions, as well as in $y$-direction in the absence of the pinching 
wall. The size of the simulation box is $16a$ in all directions (particle radius $a=5$ lattice units). We vary the aperture of 
the pinching gap $D$ by translating the obstacle center along the $y$ direction. The initial position of the particle, upstream 
to the obstacle ($x\approx - 7 a$), is given by the initial offset $b_{in}$, and the final offset is calculated at the 
symmetric position with respect to the $y$-axis. Here we note that the code implementing the lattice Boltzmann method is designed 
to construct a channel, such that two walls are simultaneously added on the opposite faces of the simulation box (for example, 
see the top and bottom walls in figure \ref{fig:systemGeometry}(a)). However, preliminary simulation tests showed that in a 
sufficiently large box (i.e., $16 a$ in $x-,~y-$ and $z$-directions in this work), the effect of the wall opposite to the pinching 
wall is negligible.

The parameter space for our investigation is characterized by three dimensionless numbers: the Reynolds number $Re$, 
quantifying fluid inertia, the Stokes number $St$, quantifying particle inertia, and the ratio of the particle diameter 
to the aperture of the pinching gap $d/D$, quantifying the extent of pinching. We define the Reynolds and Stokes numbers 
based on the bulk (Stokes) settling velocity and the radius of the particle, 
\begin{align}
Re &= \frac{a U_{Stokes}}{\nu} = \frac{a}{\nu}\left(\frac{2}{9}\cdot\frac{\rho_p}{\rho_f}\cdot\frac{a^2 g}{\nu}\right),\label{eqn:defRe}\\
St = \frac{\tau_p}{\tau_f} &= \frac{2}{9}\cdot\frac{\rho_p}{\rho_f} Re = a g \left(\frac{2}{9}\cdot\frac{\rho_p}{\rho_f}\cdot\frac{a}{\nu}\right)^2,\label{eqn:defSt}
\end{align}
where $\tau_p = m/6\pi\mu a$ is the inertial relaxation time for an isolated suspended particle,\cite{subramanian2006} 
$\tau_f = a/U_{Stokes}$ is the characteristic time of the flow around the particle, $\nu$ is the kinematic viscosity of 
the fluid ($\nu = 1/6$ in lattice units throughout this work\footnote{It has been reported that using kinematic viscosity 
$\nu = 1/6$ (in lattice units) hastens the simulations as well as reduces numerical errors [Ladd, A. J. C. and Verberg, R. 
(2001), J Stat Phys, Vol.104, Nos. 5/6, pp. 1191--1251], providing a basis for our choice. Also, we note that the accuracy 
of the lattice-Boltzmann method based on the inherent compressibility, is reported to be $O(Ma^2)$ in the literature [e.g., 
Ladd, A. J. C. and Verberg, R. (2001), J Stat Phys, Vol.104, Nos. 5/6, pp. 1191-1251 and Hazi, G. (2003), Phys Rev E, Vol. 
67, article 056705], where $Ma$ is the Mach number associated with the system. For our system, the largest instantaneous 
particle velocities observed in the simulations are $O(10^{-2}~\text{lattice units})$. Given that the LB-code is D3Q19, 
velocity of sound is given by $c_s^2 = 1/3$ in lattice units. Thus, $Ma \sim O(10^{-2})$ and errors related to inherent 
compressibility of the method are $O(10^{-4})$.}), $g$ is the acceleration due to the constant external force 
acting on the particle,\footnote{We note that, in our simulations we specify the constant external acceleration acting on 
the particle, which is multiplied by the particle mass to compute the net force acting on it. Therefore, the expression for 
the Stokes settling velocity in equation \ref{eqn:defRe} contains just the particle density $\rho_p$, and not the difference between 
the particle and fluid densities, $\Delta\rho = \rho_p-\rho_f$.} and $\rho_p$, $\rho_f$ are the densities of the particle and the fluid, 
respectively. We cover a range of $Re$ and $St$ values from negligible ($O(10^{-4})$-$O(10^{-3})$) to appreciable ($O(0.1)$-$O(1)$) 
by varying $\rho_p/\rho_f$ and $g$, and use four values of the $d/D$ ratio: $0.588, 0.5, 0.4$ and $0.333$. 

In general, the trajectories followed by the particles for different initial offsets $b_{in}$, and thus the minimum dimensionless 
separations $\xi_{min}$, are functions of the three dimensionless numbers: $Re,~St,~d/D$. We estimate the effect of short-range 
non-hydrodynamic interactions on a particle trajectory by comparing the range of the interactions with the corresponding minimum 
separation attained by the particle along that trajectory. Further, the particle trajectories in the Stokes regime are 
symmetric about the $y$-axis in the absence of non-hydrodynamic interactions. Therefore, the symmetry (or equivalently, asymmetry) 
exhibited by a particle trajectory also serves a diagnostic purpose: any asymmetry present in the trajectories obtained in our 
simulations is indicative of deviations from the Stokes regime (i.e., significant inertia)\citep{balvin2009, frechette2009}. The 
presence of such asymmetry can be determined by comparing the 
initial and final offsets.  Thus, in the next sections, we use the minimum separation ($\xi_{min}$) as well as the offsets ($b_{in}$ 
and $b_{out}$) as representative variables to investigate the effect of non-hydrodynamic interactions, inertia 
and pinching on the particle trajectories. Throughout this article, for brevity and clarity, we present results corresponding to 
$b_{in} = 2 a$. In appendix \ref{sec:appA}, we present the minimum separation $\xi_{min}$ for the entire range of $b_{in}$ considered ($1.6a$--$2.2a$). 

\section{Hard-core model for non-hydrodynamic interactions}\label{sec:onlyWall}
\subsection{Stokes regime and the critical offset}\label{subsec:hardWallModel}
We have shown in previous work that the minimum separation $\xi_{min}$ decreases dramatically with the initial offset $b_{in}$, 
especially when $b_{in}$ becomes comparable to the obstacle radius (or, the mean of particle and obstacle radii for unequal 
sizes).\citep{frechette2009,risbud2013} An initial offset of $b_{in}\sim a$, for example, results in a minimum separation 
$\xi_{min}\sim O(10^{-3})$. Therefore, for a particle with its radius in the millimeter range, the corresponding minimum separation 
amounts to a few microns, while for micrometer-sized particles, the predicted minimum separations are at atomic length-scales. In 
reality, micrometer-sized particles exhibit various non-hydrodynamic interactions at separations $O(100$ nm$)$, of which we are concerned with those of a repulsive nature in the context of this work, such as, 
electrostatic double layer repulsion,\citep{russel1989} solid-solid contact due to surface roughness\citep{smart1989, smart1993} and 
steric repulsion.\citep{russel1980} Following previous studies, we use a simple model that describes these interactions with a hard-core  
repulsive potential.\citep{frechette2009, rampall1997, brady1997, dacunha1996, davis1992} In this approximation, if the dimensionless 
range of the repulsive interaction is $\epsilon$, any surface-to-surface separation $\xi<\epsilon$ is not possible, since it is inside 
the core of the interaction. Consequently, if a particle trajectory reaches a 
minimum separation $\xi_{min}$ in the absence of non-hydrodynamic interactions, by comparing it with $\epsilon$ 
we investigate the effect of these interactions. Moreover, we assume that the hard-core repulsion is the only effect of the 
potential, in that, the hydrodynamic mobility of the suspended particle is not affected.\citep{frechette2009, 
rampall1997, brady1997, dacunha1996, davis1992}

\begin{figure}
\includegraphics[width=1.1\fw]{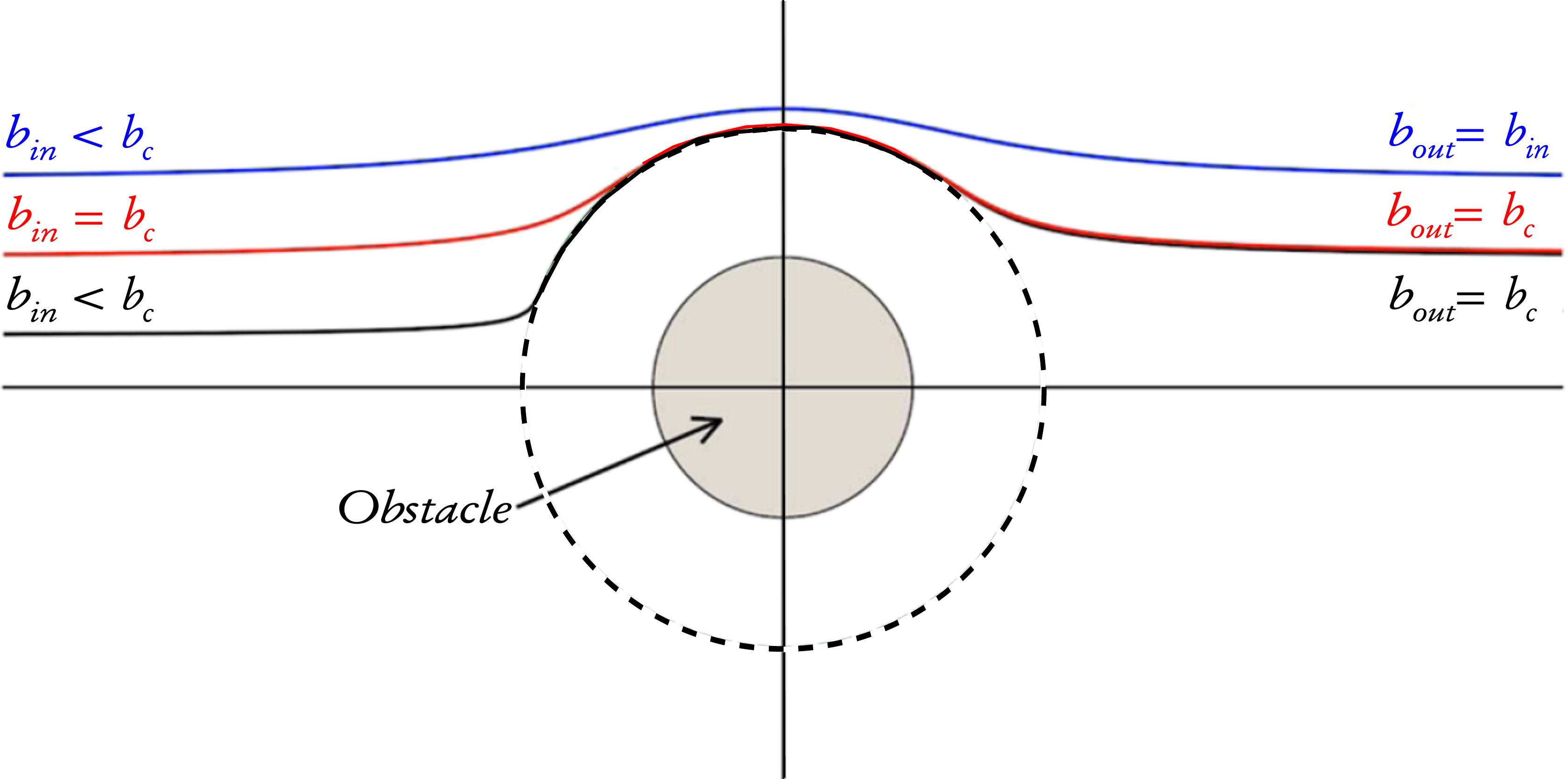}
\caption{Trajectories corresponding to the three possible types of particle-obstacle `collisions', adapted from 
\cite{balvin2009}. The particle travels from left (incoming half of the trajectory) to right (outgoing 
half of the trajectory). The dashed circle represents the excluded volume inaccessible to the particle centers.}
\label{fig:threeTypesOfCollisions}
\end{figure}

In the absence of inertia, there is a one-to-one relationship between the minimum separation and the initial offset.
\citep{frechette2009,risbud2013} We can thus define a {\it critical} offset $b_c$, as the final offset resulting from a 
minimum separation equal 
to the range of the repulsive interaction i.e., $\xi_{min} = \epsilon$ (see figure \ref{fig:threeTypesOfCollisions}). In previous work, 
we have established that the critical offset $b_c$ 
completely characterizes the asymmetry in the particle trajectory caused by the repulsive interaction.\citep{frechette2009,balvin2009} Any trajectory beginning with an initial offset smaller than the critical value ($b_{in}<b_c$) would result in a minimum separation 
smaller than the range of the repulsive interaction. Since such separations are unattainable in the presence of the interaction, 
once the particle attains $\epsilon$, the separation remains constant and equal to $\epsilon$ on the approaching side and, as shown in 
figure \ref{fig:threeTypesOfCollisions}, the outgoing half of the trajectory collapses onto the trajectory with $b_{out}=b_c$. 
On the other hand, for $b_{in}\ge b_c$, the trajectories, unaffected by non-hydrodynamic interactions, are symmetric in the absence of 
inertia. The single-valued function relating $\xi_{min}$ and $b_{in}$ (or $b_{out}$) is the same as that relating $\epsilon$ and $b_c$.
\cite{frechette2009,risbud2013} Therefore, we can predict the dependence of $b_c$ on the pinching gap for a given $\epsilon$ by 
computing $\xi_{min}$ as a function of $b_{in}$ for different $d/D$ ratios. {\em In conclusion, the critical offset is a macroscopic 
and experimentally measurable variable, which serves as a probe/proxy for the range of the repulsive interaction, a microscopic 
effective variable that is experimentally less accessible.}\cite{risbud2013}

\subsection{Inertia effects on the critical offset}\label{subsec:hardWallModelInertia}
\begin{figure}
\includegraphics[width=1.5\fw]{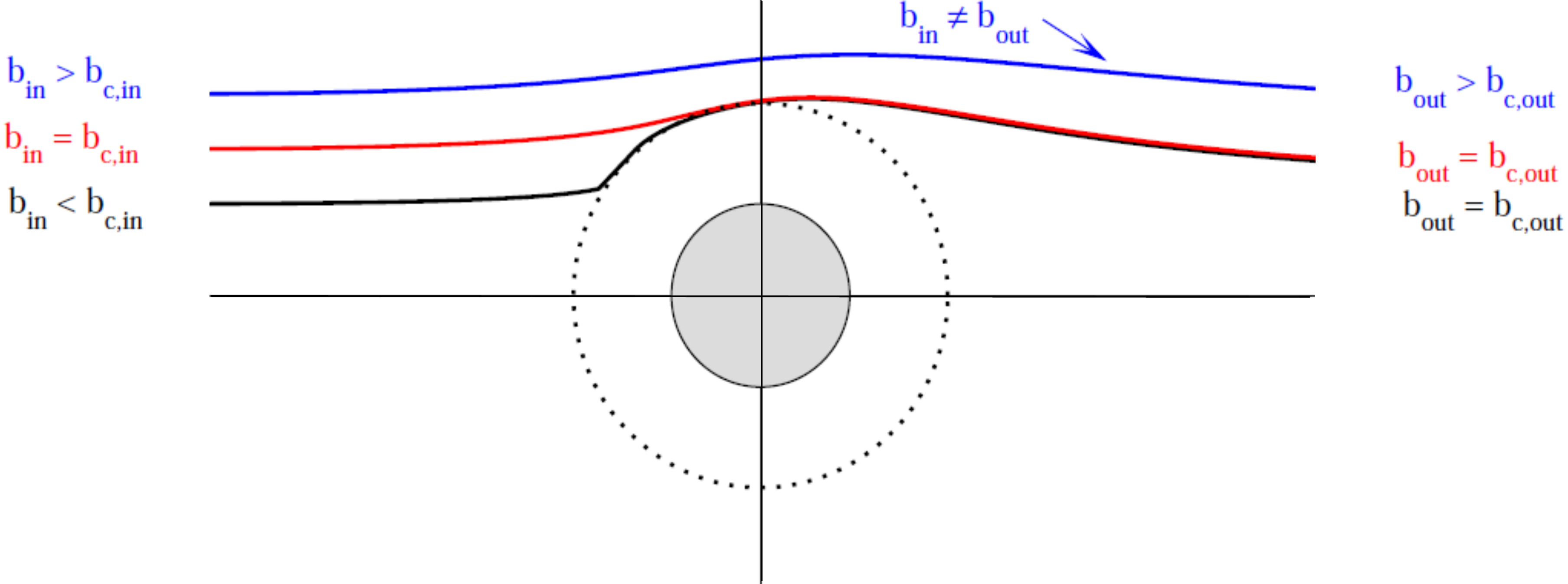}
\caption{In presence of significant inertia (in this particular case, particle inertia, $St = 1.11$), the trajectories are asymmetric 
even in the absence of non-hydrodynamic interactions. Such asymmetry leads to two critical offsets, viz.- $b_{c,in}, b_{c,out}$ as shown.}
\label{fig:threeTypesOfCollisionsInertia}
\end{figure}

The fore-aft symmetry of the trajectories is broken in the presence of inertia (see figure \ref{fig:threeTypesOfCollisionsInertia}). 
Therefore, in the presence of non-hydrodynamic interactions, we need to define two distinct critical offsets: the critical initial-offset 
$b_{c,in}$ and the critical final-offset $b_{c,out}$, as shown in figure \ref{fig:threeTypesOfCollisionsInertia}.
The critical initial-offset $b_{c,in}$ is defined in the same manner as in case of negligible inertia. The definition 
of the critical final-offset ($b_{c,out}$) needs the auxilliary assumption that inertia is negligible when the particle approaches 
small surface-to-surface separations from the obstacle. Under this assumption, the final offset would also present a critical behavior 
such that, all trajectories with $b_{in}<b_{c,in}$ would collapse onto a single trajectory downstream to the obstacle, which in turn 
defines the critical final-offset $b_{c,out}$ corresponding to $b_{c,in}$.

\section{Results and implications of the hard-core model}\label{sec:bothInertiaNdWall}
\subsection{Effect of pinching in the Stokes regime}\label{subsec:applModelOnlyWall}
\begin{figure}
\includegraphics[width=1.5\fw]{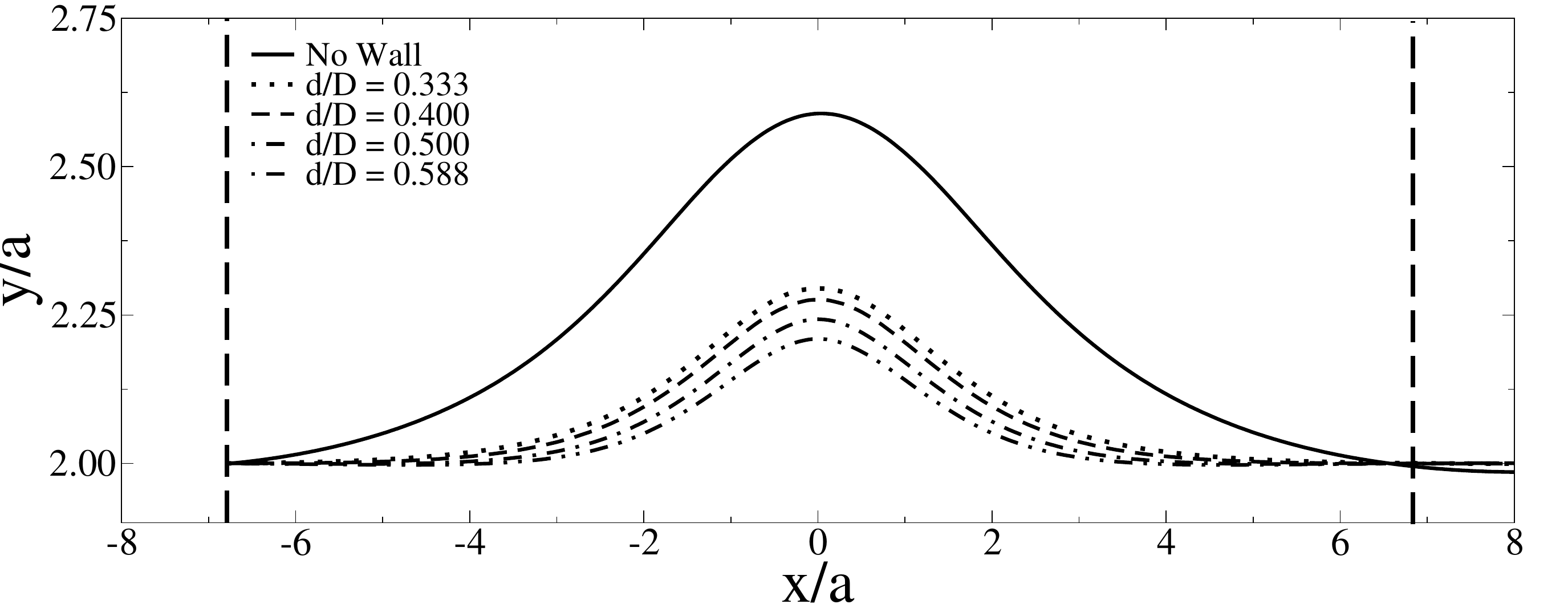}
\caption{Various trajectories with $b_{in}=2.0$, exhibiting the effect of pinching in the Stokes regime 
($Re = 5\times10^{-4}$ and $St = 1.11\times10^{-4}$). The vertical 
dashed lines at $x\approx\pm7a$ represent the initial and final $x$-coordinates of the particle, respectively.} 
\label{fig:sampleTrajectoriesWall}
\end{figure}
In figure \ref{fig:sampleTrajectoriesWall}, we show representative trajectories corresponding to different apertures of the pinching gap, 
for the same initial offset $b_{in} = 2 a$ and with negligible fluid and particle inertia. As expected, the trajectories remain fore-aft 
symmetric in all cases. When the aperture of the pinching gap is reduced, we observe that the particle follows trajectories that 
get significantly closer to the obstacle. 

In light of the hard-core model discussed before, let us now consider the effect of the non-hydrodynamic interactions on the trajectories 
through a pinching gap in the Stokes regime. Consider trajectories with the initial offset $b_{in}=2 a$ (see figure 
\ref{fig:sampleTrajectoriesWall}), and non-hydrodynamic interactions with a dimensionless range $\epsilon=\epsilon_0$. Let the critical 
offset corresponding to $\epsilon_0$ in the absence of pinching be $b_c=b_{0}<2 a$. Therefore, in the absence 
of pinching the particle trajectories under consideration attain a minimum separation $\xi_{min}>\epsilon_0$, and the 
non-hydrodynamic 
interactions have no effect on them (e.g., the top-most trajectory in figure \ref{fig:threeTypesOfCollisions}). As the aperture of the 
pinching gap is reduced, while keeping the initial offset constant at $2 a$, figure \ref{fig:sampleTrajectoriesWall} shows that the particle comes closer 
to the obstacle. Eventually, at a particular aperture, we get $\xi_{min}=\epsilon_0$, which implies that the critical offset has increased 
from $b_{0}$ to $2 a$. In other words, {\em the critical offset that leads to a given minimum separation increases as the aperture of the  
pinching gap decreases}. This result is in qualitative agreement with our experiments.\cite{luo2011} In appendix \ref{sec:appA} 
we arrive at the same inference directly by considering the minimum separation $\xi_{min}$ as a 
function of the initial offset $b_{in}$ (see figures \ref{fig:onlyWallNoInertia} and \ref{fig:onlyInertiaNoWall}, and the corresponding discussion). 

\subsection{Effect of inertia in the absence of pinching}\label{subsec:applModelOnlyInertia}
\begin{figure}
\includegraphics[width=1.5\fw]{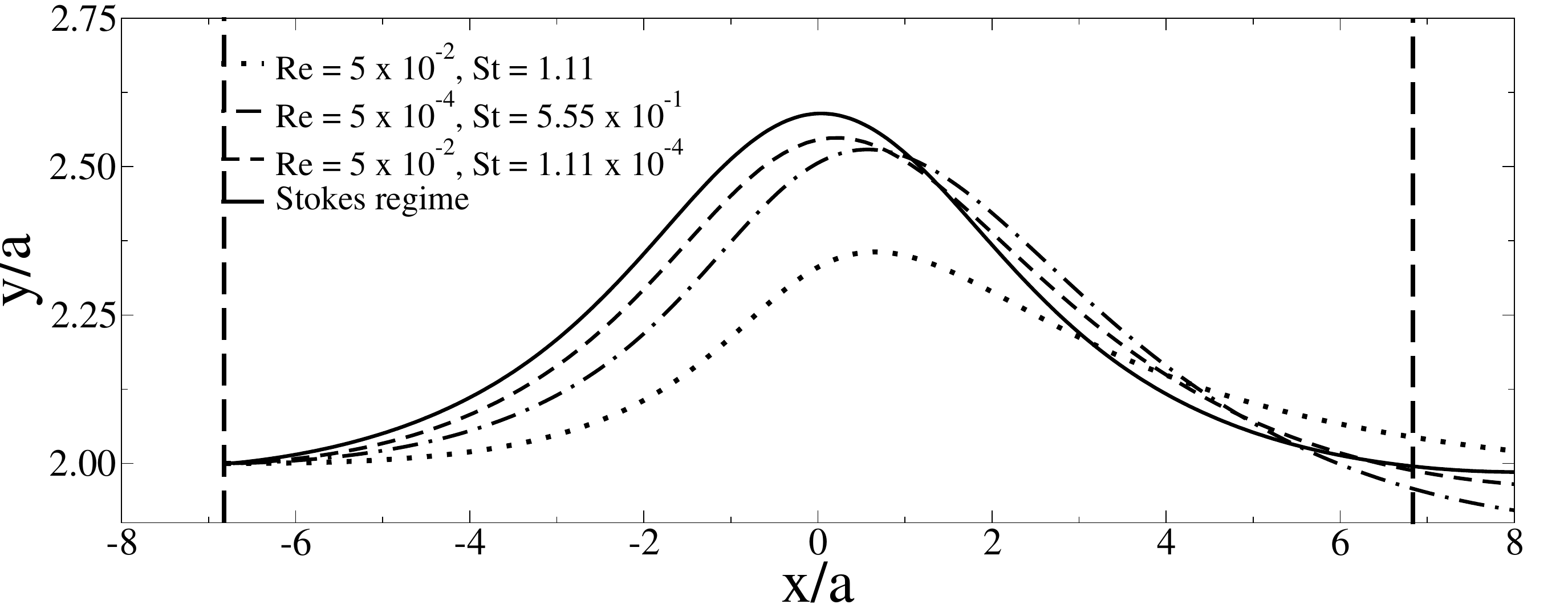}
\caption{Various trajectories with $b_{in}=2 a$, exhibiting the effect of inertia in the absence of pinching. 
The vertical dashed lines at $x\approx\pm7 a$ represent the initial and 
final $x$-coordinates of the particle, respectively.} 
\label{fig:sampleTrajectoriesInertia}
\end{figure}

\begin{figure}
\includegraphics[width=\fw]{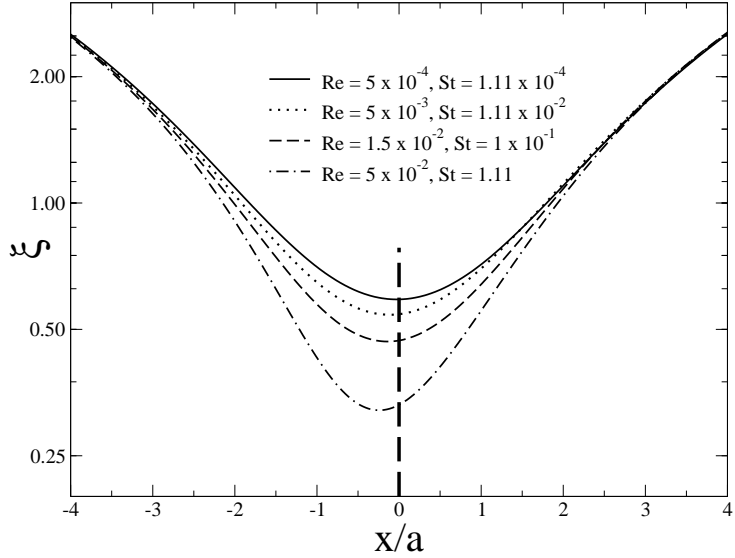}
\caption{The dimensionless surface-to-surface separation between the particle and the obstacle, as a function of the dimensionless 
$x$-coordinate of the particle. The plot shows the combined effect of particle and fluid inertia on the dimensionless separation along 
particle trajectories for $b_{in}=2 a$, in the absence of pinching. The vertical dashed line depicts the symmetry plane of the system 
at $x=0$.}
\label{fig:ximinVsXbothReSt}
\end{figure}

Figure \ref{fig:sampleTrajectoriesInertia} shows representative trajectories (with $b_{in}=2 a$) in the presence of inertia effects, but 
in the absence of a pinching wall. We observe that the effect of inertia is similar to that of pinching discussed earlier, such that the 
particle gets closer to the obstacle as the magnitude of (fluid and/or particle) inertia increases. Although the effects of pinching and 
inertia on the minimum separation are similar, in the presence of inertia the minimum separation is no longer attained at the symmetry 
plane ($x=0$) but upstream to the obstacle-center (at negative $x$-values, see figure \ref{fig:ximinVsXbothReSt}). We also observe that 
the $x$-coordinate at which the minimum separation is attained decreases with increasing inertia (similar plots showing the individual 
effects of fluid and particle inertia can be found in appendix \ref{sec:appA}).

\subsection{Effect of pinching and inertia}\label{subsec:applModelBoth}
\begin{figure}
\includegraphics[width=\fw]{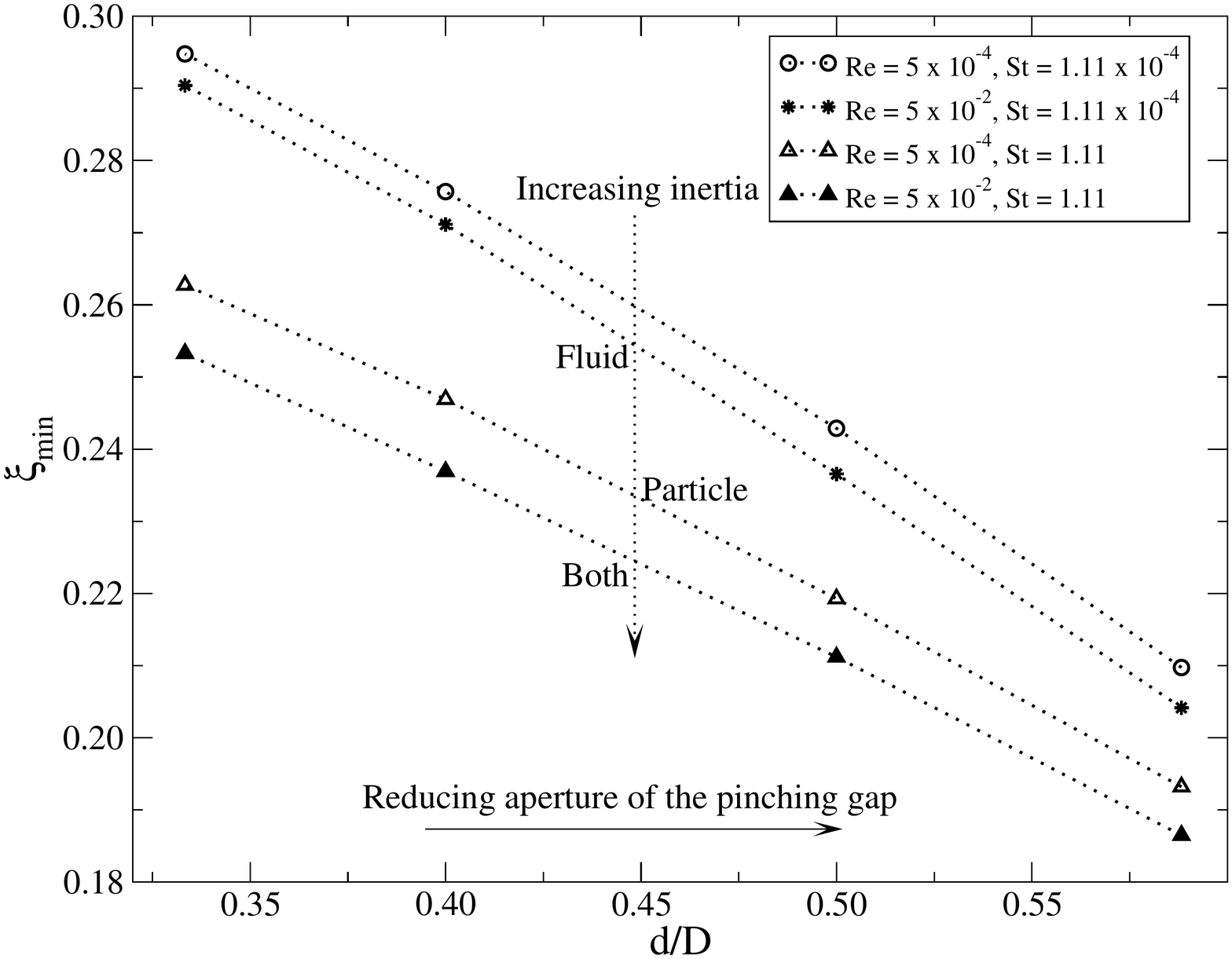}
\caption{Effect of reducing the aperture the of pinching gap ($d/D\uparrow$) on the minimum 
surface-to-surface separation between 
the particle and the obstacle, for different magnitudes of inertia and a fixed initial offset, $b_{in} = 2 a$.} 
\label{fig:ximindOverDBin2AllInertia}
\end{figure}
In figure \ref{fig:ximindOverDBin2AllInertia}, we consolidate our results and present the minimum separation $\xi_{min}$ 
for all the cases of pinching and inertia considered in this work, at $b_{in}=2 a$. 
As discussed so far, we observe that reducing the aperture of the pinching gap causes the particle to attain a smaller 
minimum surface-to-surface separation in all the cases shown, with or without particle and/or fluid inertia. Further, the 
figure also shows that for any fixed aperture of the pinching gap the minimum separation attained by the particle 
decreases with increasing inertia. 
Our simulations also indicate that the effect of particle inertia on $\xi_{min}$ is attenuated in the presence of the pinching
 wall. The attenuation could be 
due to the hydrodynamic resistance offered by the pinching wall, which reduces the particle velocity, thereby diminishing 
the effective magnitude of particle inertia (note that for calculating the Stokes and Reynolds numbers reported in this work, we have used 
the bulk settling velocity of spheres). The detailed plots of $\xi_{min}$ as a function of $b_{in}$ for different Stokes and 
Reynolds numbers are presented in figures \ref{fig:onlyInertiaNoWall} and \ref{fig:bothReStNoWall}, and the effect of pinching on 
these plots is shown in figures \ref{fig:bothInertiaNdWallat30}--\ref{fig:bothInertiaNdWall}, in appendix \ref{sec:appA}. 

\begin{figure}
\includegraphics[width=\fw]{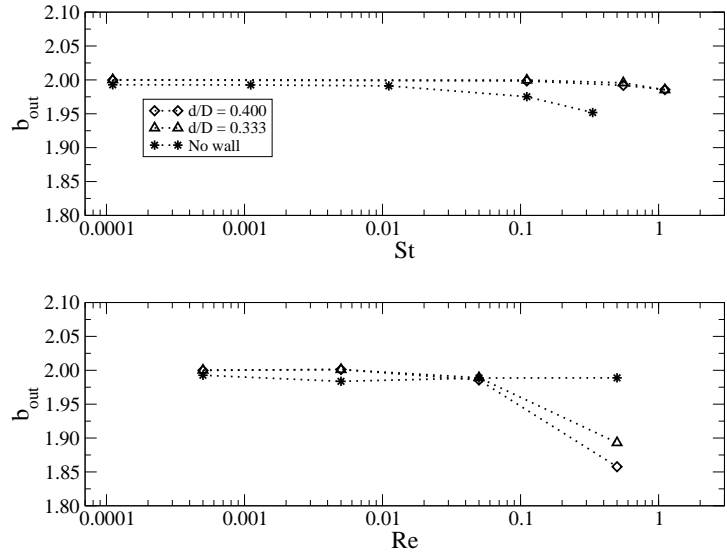}
\caption{The final offset $b_{out}$ (corresponding to $b_{in}=2a$) as a function of Stokes (top) and Reynolds (bottom) numbers, for different 
apertures of the pinching gap.} 
\label{figbout:onlyInertiaNoWall}
\end{figure}

\begin{figure}
\includegraphics[width=\fw]{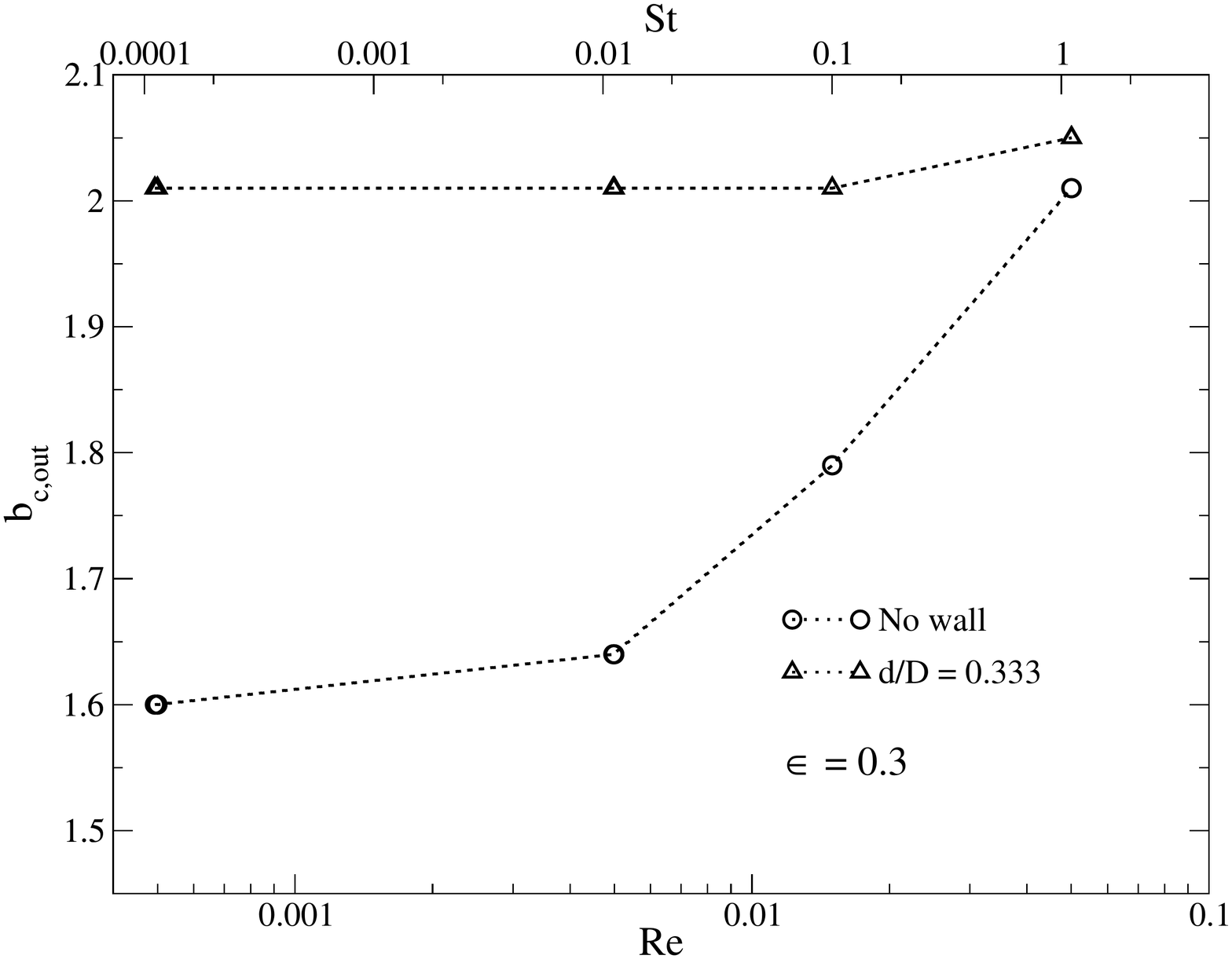}
\caption{The critical final-offset $b_{c,out}$ corresponding to $\epsilon = 0.3$, as a function of Reynolds and Stokes numbers, 
in the presence and the absence of a pinching wall.} 
\label{fig:bcoutVsRhoRatio}
\end{figure}

We are also interested in the effect of inertia in combination with pinching on the final offset $b_{out}$, since changes in the final 
offset as a function of inertia could lead to separation based, for example, on density differences. In figure \ref{figbout:onlyInertiaNoWall}, 
we present the effects of particle and fluid inertia on the final offset, for $b_{in}=2 a$. In terms of particle inertia, we observe 
that the final offset decreases with increasing Stokes number. The final offset decreases less when the trajectories undergo pinching, 
than in the absence of pinching. As mentioned earlier, we suspect that this behavior is due to the diminished particle inertia resulting from the 
hydrodynamic resistance offered by the wall. In the case of fluid inertia, we observe that for $Re<0.1$ the final offset is not significantly affected, 
irrespective of the presence or absence of the pinching wall. For higher magnitudes of fluid inertia ($Re\sim O(1)$), we observe that 
the final offset decreases with the aperture of the pinching gap. Such decrease is comparable to the effect of lift forces in the proximity 
of the pinching wall, which render the final offset ill-defined. In appendix \ref{sec:appA}, 
we present the minimum separation $\xi_{min}$ as a function of the initial offset $b_{in}$, which is consistent 
with the attenuation of particle inertia and the presence of lift forces (figures \ref{fig:bothInertiaNdWallat30}, 
\ref{fig:bothInertiaNdWallat17} and the accompanying discussion).

Figure \ref{fig:ximindOverDBin2AllInertia} shows that the minimum 
separation decreases for smaller apertures of the pinching gap for a constant magnitude of inertia. Therefore, similar to the discussion in 
\S\ref{subsec:applModelOnlyWall}, we can infer that the critical initial-offset $b_{c,in}$ increases with pinching. Further, the 
minimum separation also decreases with an increasing magnitude of inertia, for a fixed aperture of the pinching gap. Thus, the critical 
initial-offset increases with increasing particle inertia (see figures \ref{fig:onlyInertiaNoWall} and 
\ref{fig:bothInertiaNdWallat30}--\ref{fig:bothInertiaNdWall} in appendix \ref{sec:appA}).
Since $b_{c,out}$ is a monotonically increasing function of $b_{c,in}$, one might expect that $b_{c,out}$ also increases 
with pinching and inertia. On the other hand, figure \ref{figbout:onlyInertiaNoWall} shows that for a constant $b_{in}$, the final offset decreases due 
to the effect of inertia. Due to such competing effects, the behavior of the critical final-offset as a function of the aperture 
of the pinching gap and inertia is not obvious. Our simulations indicate, as shown in figure \ref{fig:bcoutVsRhoRatio}, that the critical 
final-offset $b_{c,out}$ indeed increases with inertia, but the change becomes substantially smaller in the presence of pinching, provided that 
the effect of lift forces generated due to the pinching wall are negligible (see also $\xi_{min}$ as a function of $b_{out}$ in figures 
\ref{figOut:bothInertiaNdWallat30} and \ref{figOut:bothInertiaNdWallat17} in appendix \ref{sec:appA}).

Thus, in conclusion, {\em both critical offsets increase with inertia and/or pinching, for a constant range 
of the non-hydrodynamic interactions $\xi_{min} = \epsilon$}. Thus, in a mixture of particles of the same size, 
a particle with sufficiently higher 
density would exhibit larger critical final-offset $b_{c,out}$ than a particle with lower density, thereby rendering 
its separation possible (assuming that both particles are subjected to non-hydrodynamic interactions of similar range). In the presence of a 
pinching wall, a similar inference can be drawn. Although, the change in the critical offsets due to inertia can be expected to be small 
because of the attenuation of inertia effects by the wall. 

\section{Summary}\label{sec:summary}
We have shown that in the presence of inertia effects and/or a pinching wall, suspended particles driven around an obstacle by 
a constant force, get closer to the obstacle. 
Using a simple hard-core repulsion model for non-hydrodynamic interactions (such as solid-solid contact due to surface roughness) our 
observations lead to the definitions of critical initial- and final-offsets (both offsets are equal when inertia is negligible). The offsets 
increase upon decreasing the aperture of the pinching gap and/or increasing inertia.
When the inertia and pinching effects are combined 
together, the final offset varies only weakly with inertia. In this case, we surmise that the presence of a pinching wall attenuates the effect of particle 
inertia. 
Therefore, the presence of a geometric constriction might, in some case, hinder inertia-based separations.

We thank Prof. A. J. C. Ladd for making the LB code -- {\em Susp3d} -- available to us. This work is partially supported by the 
National Science Foundation Grant Nos. CBET-0933605, CMMI-0748094 and CBET-0954840. This work used the 
resources of the National Energy Research Scientific Computing Center, which is supported by the Office of Science of the U.S. Department 
of Energy under Contract No. DE-AC02-05CH11231.

\appendix

\section{An alternate perspective: the minimum separation as a function of the initial offset}\label{sec:appA}
Here, we present the results of our simulations in a different manner than the main text by considering the minimum separation 
$\xi_{min}$ as a function of the initial or final offsets.  
First, we show the effect of decreasing the aperture of the pinching gap for negligible fluid and particle inertia (the Stokes 
regime, figure \ref{fig:onlyWallNoInertia}). Then we show the effect of inertia, in the absence of a pinching wall (figures 
\ref{fig:onlyInertiaNoWall} and \ref{fig:bothReStNoWall}). Finally we present the cases when fluid and/or particle 
inertia is significant, in the presence of a pinching wall (figures \ref{fig:bothInertiaNdWallat30}, \ref{fig:bothInertiaNdWallat17} 
and \ref{fig:bothInertiaNdWall}). As discussed in the main text, the relationship between $\xi_{min}$ and $b_{in}$ 
shown here, is the same as that between the range of the repulsive non-hydrodynamic 
interactions and the critical offset.\cite{frechette2009,risbud2013} Therefore, representing the data in terms of 
$\xi_{min}$--$b_{in}$ plots provides additional clarity to the discussion concerning the critical offset provided 
in the main text.

\subsection{Pinching gap, inertia and minimum separation}\label{sec:results}
\begin{figure}
\includegraphics[width=\fw]{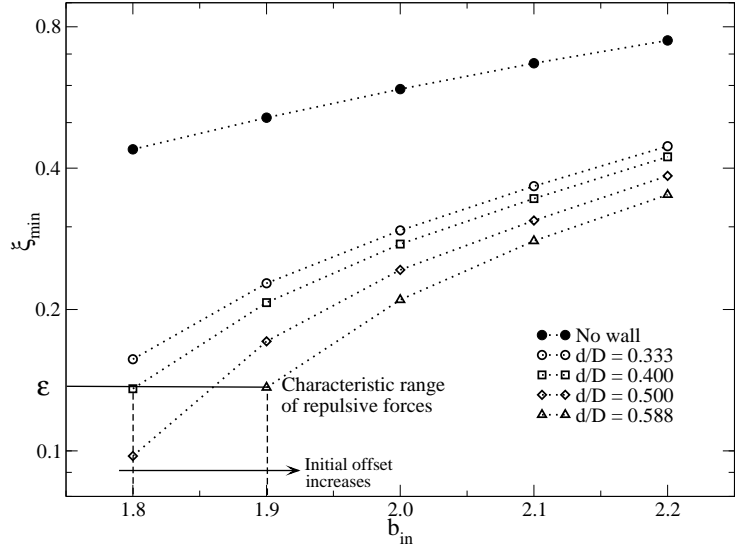}%{figureA1.pdf}%{../figures/ximinBin_onlyD_12032012.pdf}
\caption{The minimum separation $\xi_{min}$ as a function of the initial offset $b_{in}$, $Re$ and $St$ are negligible.}
\label{fig:onlyWallNoInertia}
\end{figure}
\begin{figure}
\includegraphics[width=\fw]{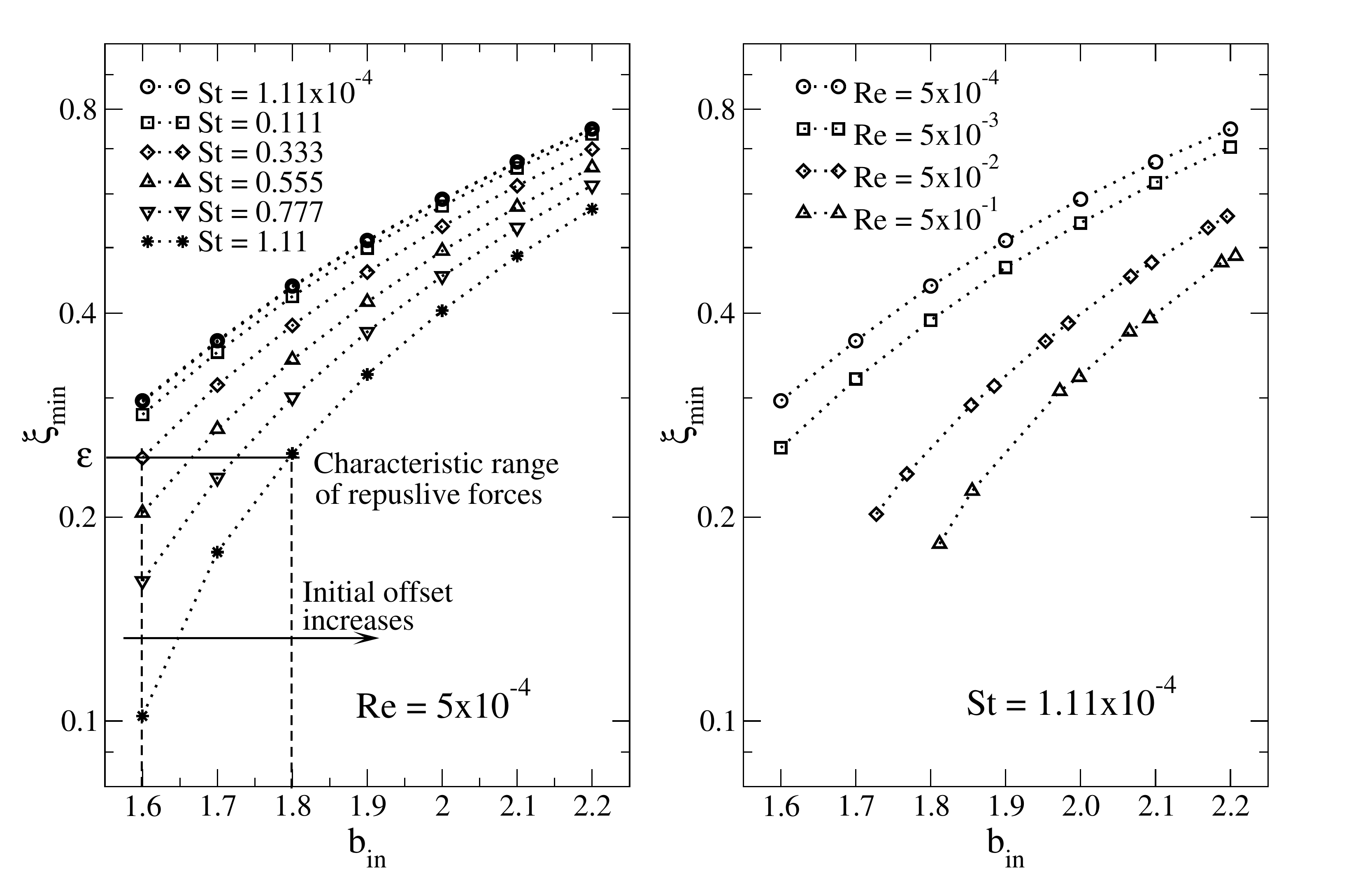}%{figureA2.pdf}%{../figures/ximinBin_Re-St_twoGraphs.pdf}
\caption{The minimum separation $\xi_{min}$ as a function of the initial offset $b_{in}$, in the absence of pinching. 
(Left) effect of particle inertia when fluid inertia is negligible, (right) effect of fluid inertia when particle 
inertia is negligible.}
\label{fig:onlyInertiaNoWall}
\end{figure}
\begin{figure}
\includegraphics[width=\fw]{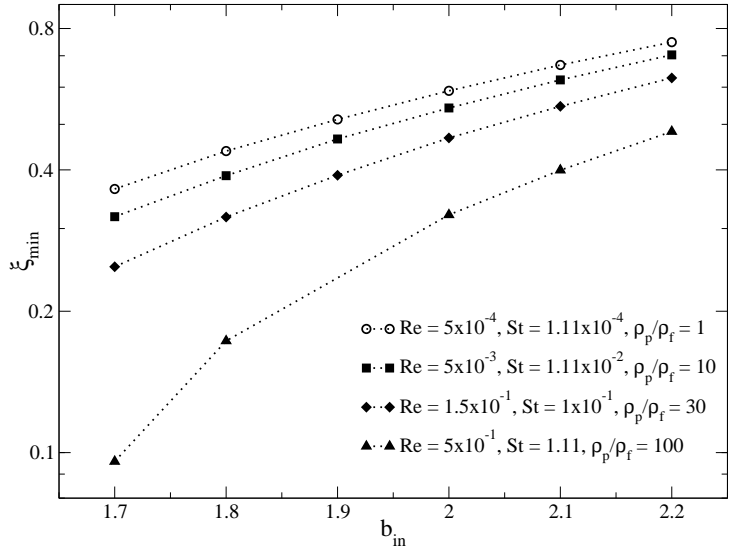}%{figureA3.pdf}%{../figures/ximinBin_bothReSt_28032012.pdf}
\caption{The minimum separation $\xi_{min}$ as a function of the initial offset $b_{in}$ showing the effect of significant fluid as well 
as particle inertia in the absence of pinching.}
\label{fig:bothReStNoWall}
\end{figure}

\begin{figure}
\includegraphics[width=\fw]{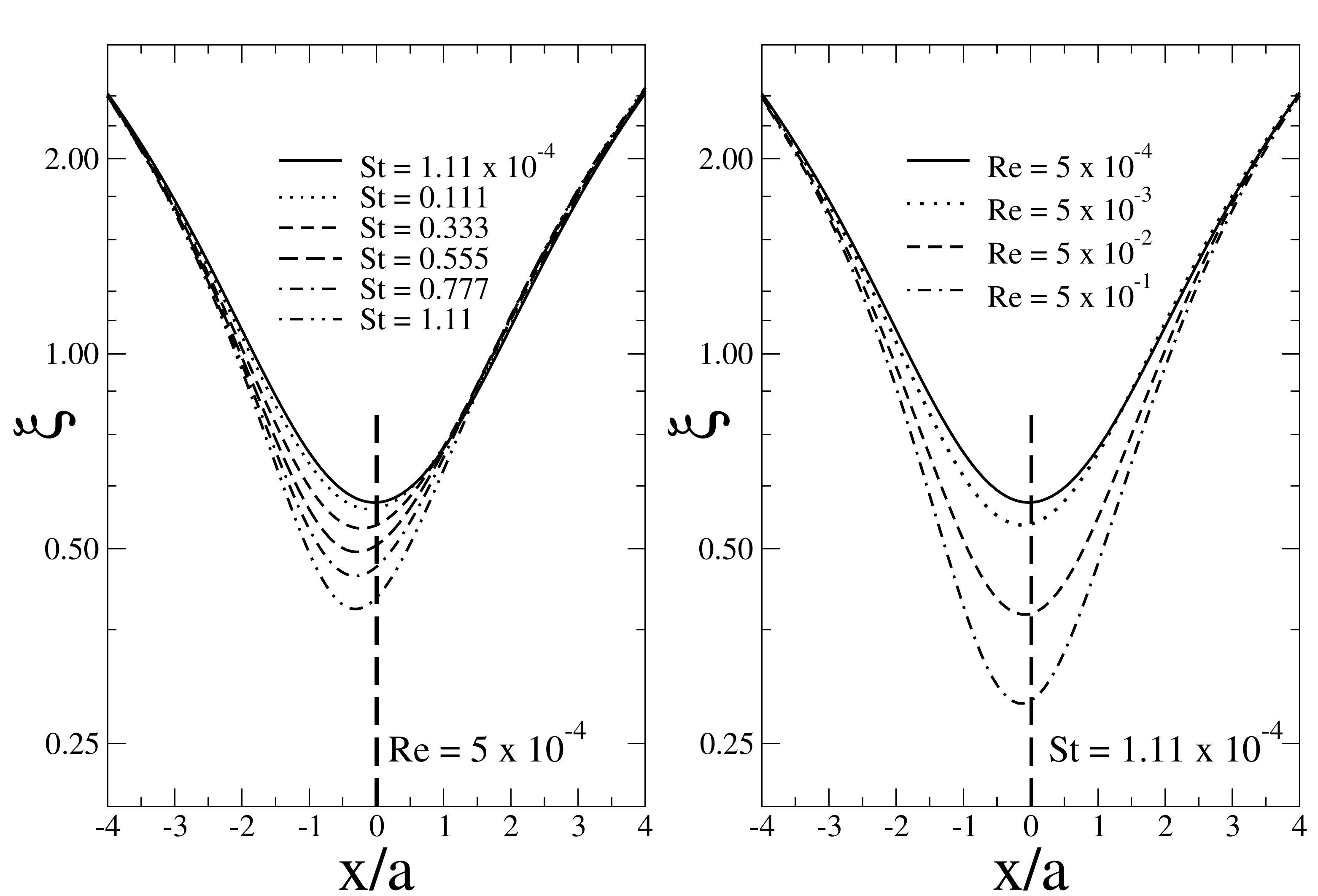}%{figureA4.pdf}%{../figures/ximinVsX_bin2ReSt-twographs_03042012.pdf}
\caption{The dimensionless surface-to-surface separation $\xi$ as a function of the dimensionless x-coordinate of the particle. 
The x-coordinate at which the minimum separation is attained along a trajectory, varies with inertia. (Left) Effect of particle inertia, 
(right) effect of fluid inertia. The minimum separation is attained before the $x=0$ symmetry-plane in both cases.}
\label{fig:ximinVsXReSttwoGraphs}
\end{figure}

In figure \ref{fig:onlyWallNoInertia}, we present the effect of decreasing the aperture of the pinching gap on the minimum 
separation $\xi_{min}$, in the Stokes regime. As noted in the main text, for a given initial offset we observe that $\xi_{min}$ 
decreases in the presence of a pinching wall. This is seen by comparing the results shown in solid circles with those in open 
symbols, which correspond to the absence and presence of a pinching wall, respectively. Moreover, comparing the different 
curves with open symbols, which correspond to different apertures of the pinching gap, it is clear that decreasing the aperture 
leads to smaller minimum separations, as the curves shift downwards. Next, in figure \ref{fig:onlyInertiaNoWall} we present the 
effect of inertia in the absence of pinching. 
The two plots in figure \ref{fig:onlyInertiaNoWall} show the effects of fluid and particle inertia on the minimum 
separation $\xi_{min}$, independently. It is clear that $\xi_{min}$ decreases when the magnitude of the particle 
or fluid inertia, increases. The combined effect of particle and fluid inertia is presented in figure 
\ref{fig:bothReStNoWall}, where we report results corresponding to particle to fluid density ratios $1, 10, 30$, and $100$. 
As explained in the text, inertia causes the particle trajectories to be asymmetric. This leads to the minimum separation 
taking place before $x=0$. This is corroborated in figure \ref{fig:ximinVsXReSttwoGraphs} for fluid and particle inertia 
acting independently. It also shows that the $x$-coordinate 
at which the minimum separation is attained, decreases with increasing inertia.
\begin{figure}
\includegraphics[width=\fw]{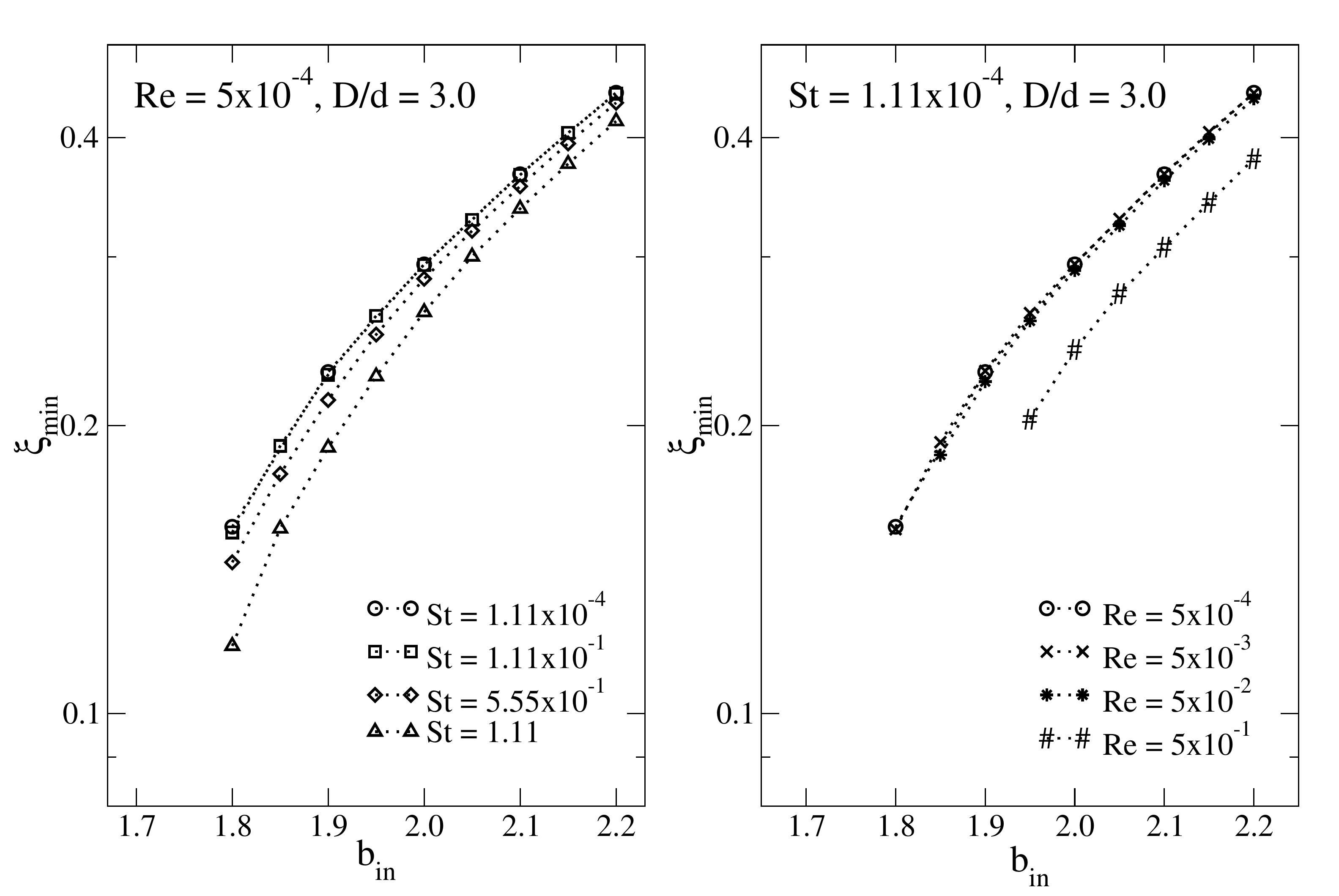}%{figureA5.pdf}%{../figures/ximinBin_Dbyd30wInertia_17032012.pdf}
\caption{The effect of inertia on the $\xi_{min}$--$b_{in}$ relationship in the presence of a pinching wall. The aperture of the 
pinching gap is $3d$ (i.e., $d/D = 0.333$). (Left) The effect of particle inertia, (right) the effect of fluid inertia.}
\label{fig:bothInertiaNdWallat30}
\end{figure}
\begin{figure}
\includegraphics[width=\fw]{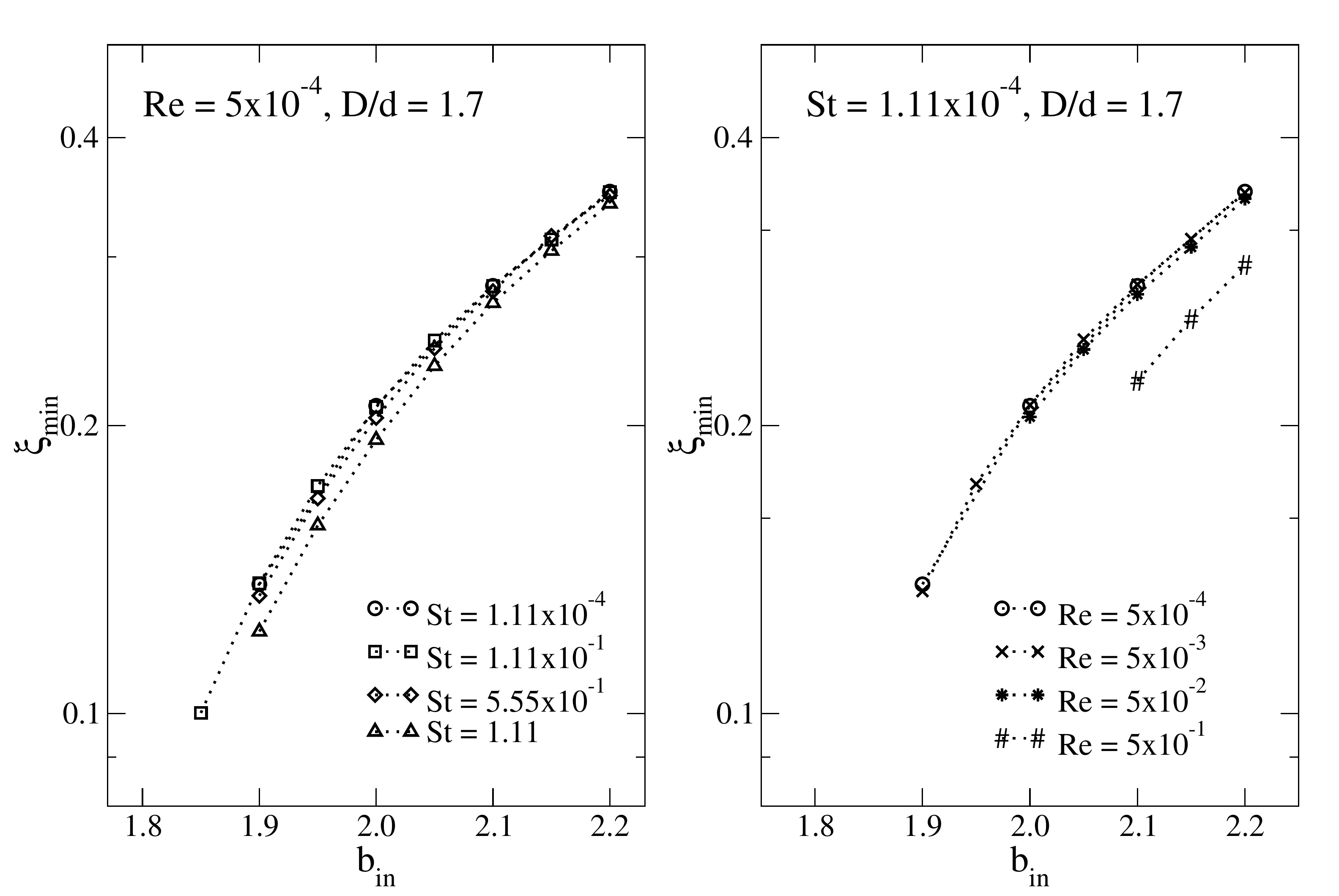}%{figureA6.pdf}%{../figures/ximinBin_Dbyd17wInertia_17032012.pdf}
\caption{The effect of inertia on the $\xi_{min}$--$b_{in}$ relationship, in the presence of pinching. The aperture of the pinching 
gap is $1.7d$ (i.e., $d/D = 0.588$). (Left) The effect of particle inertia, (right) the effect of fluid inertia.}
\label{fig:bothInertiaNdWallat17}
\end{figure}

In figures \ref{fig:bothInertiaNdWallat30} and \ref{fig:bothInertiaNdWallat17}, we present the combined effect of increasing $Re$ or 
$St$ for apertures of the pinching gap given by $D=3d$ ($d/D=0.333$) and $D=1.7d$ ($d/D=0.588$), respectively. The plot on the left 
in each figure 
shows the effect of increasing $St$ for a fixed aperture of the pinching gap and $Re$. The plot on the right in each figure shows the 
effect 
of increasing $Re$ for the same aperture of the pinching gap and constant $St$. For $b_{in}=2a$, the corresponding data is presented in 
figure $7$ of the main text. We observe that in both cases the minimum separation decreases with increasing inertia. In the case of a 
constant $Re$ (i.e., increasing $St$), we see that decreasing the aperture of the pinching gap (comparing figures \ref{fig:bothInertiaNdWallat30} 
and \ref{fig:bothInertiaNdWallat17}) attenuates the effect due to inertia of reducing  the 
minimum separation for a fixed initial offset. 
In contrast, the effect of increasing $Re$ seems to increase as the aperture of the pinching gap becomes smaller, specifically for 
$Re\sim O(1)$. In this case, however, we observe significant {\it lift} due to particle-wall interaction, which makes the discussion in 
terms of offsets less relevant. Note that for $Re \neq 0$ there are no asymptotic offsets, and we observe that for $Re\sim O(1)$ the effect 
of lift becomes significant over the length of the system. 
\begin{figure}
\includegraphics[width=\fw]{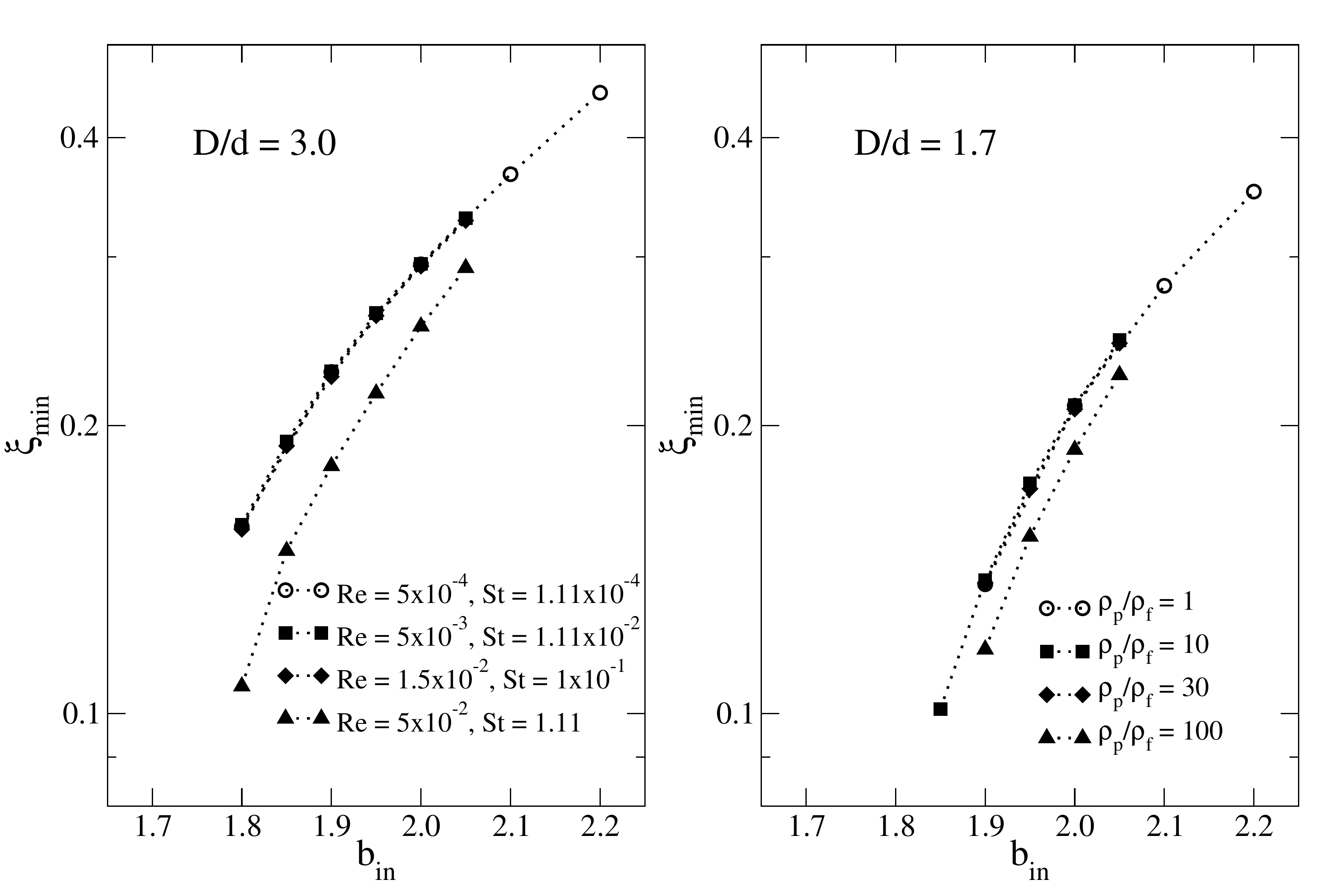}%{figureA7.pdf}%{../figures/ximinBin_bothDbyd_bothInertia.pdf}
\caption{Combined effect of the pinching wall, particle and fluid inertia on the $\xi_{min}$--$b_{in}$ relationship. (Left) the aperture of the 
pinching gap is $3d$ (i.e., $d/D=0.333$), (right) the aperture is $1.7d$ (i.e., $d/D=0.588$).}
\label{fig:bothInertiaNdWall}
\end{figure}

Finally, representative examples of the combined effect of varying all $Re$, $St$ and $d/D$ are shown in figure 
\ref{fig:bothInertiaNdWall}. The plot on the left shows the results corresponding to four pairs of $St$ and $Re$ for 
$d/D = 0.333$. These pairs correspond to the density ratios mentioned earlier, $\rho_p/\rho_f = 1,\,10,\,30,\,100$, for 
a constant force acting on the particle. The plot on the right presents the minimum separation for the same pairs of 
$Re$ and $St$ but for, $d/D = 0.588$. The results follow that, increasing inertia effects (both particle and fluid 
inertia) leads to smaller minimum separations for a fixed initial offset. Also analogous to the results discussed earlier, 
increasing the extent of pinching decreases the minimum separation corresponding to a given initial offset, but attenuates 
the relative reduction in minimum separation due to inertia effects.

\subsection{Minimum separation, critical offsets and non-hydrodynamic interactions}\label{sec:discussion}
Similar to the discussion in the main text, here we present the effect that inertia and a pinching wall have on the critical 
final-offset. It is clear from the results discussed both in the main text as well as in the previous 
section that the minimum separation $\xi_{min}$ decreases when the aperture of the pinching gap is reduced, or the magnitude of 
inertia is increased, or both. 
Using the $\xi_{min}$--$b_{in}$ plots introduced in the previous section, we elaborate more on the idea of the critical 
initial-offset $b_{c,in}$, and the 
effect of pinching and/or inertia on it. Further, extending the method, we plot $\xi_{min}$ as a function of the final offset 
$b_{out}$ for various cases of pinching and inertia, and comment on the behavior of the critical final-offset $b_{c,out}$.

\begin{figure}
\includegraphics[width=\fw]{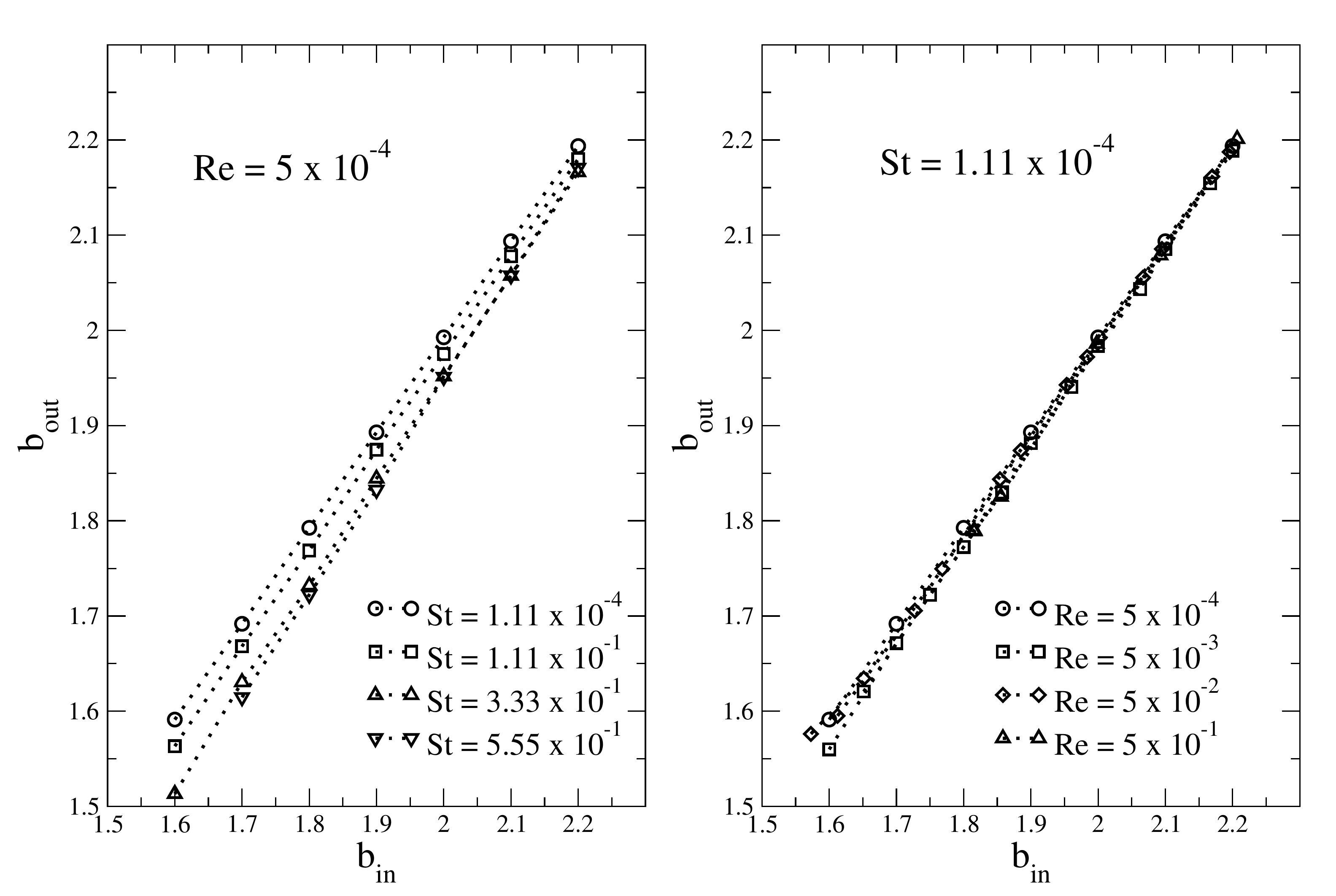}%{figureA8.pdf}%{../figures/boutBin_Re-St_twoGraphs}
\caption{The outgoing offset $b_{out}$ as a function of the incoming offset $b_{in}$. (Left) effect of particle inertia, 
(right) effect of fluid inertia.}
\label{fig:boutBinReSttwoGraphs}
\end{figure}

\begin{figure}
\includegraphics[width=\fw]{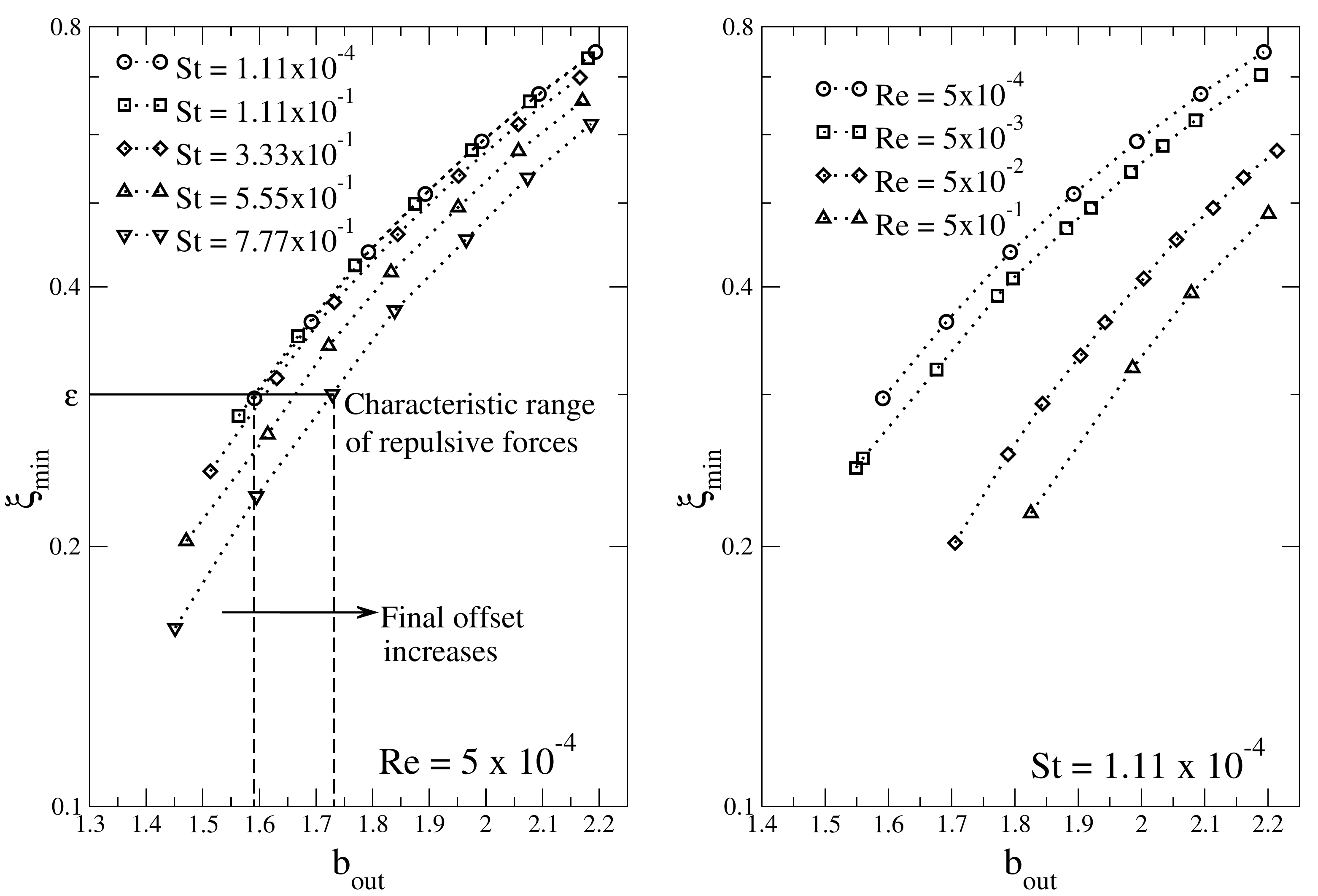}%{figureA9.pdf}%{../figures/ximinBout_Re-St_twoGraphs.pdf}
\caption{The minimum separation as a function of the final offset, (left) effect of particle inertia,
(right) effect of fluid inertia.}
\label{figbout:onlyInertiaNoWallSupp}
\end{figure}

\begin{figure}
\includegraphics[width=\fw]{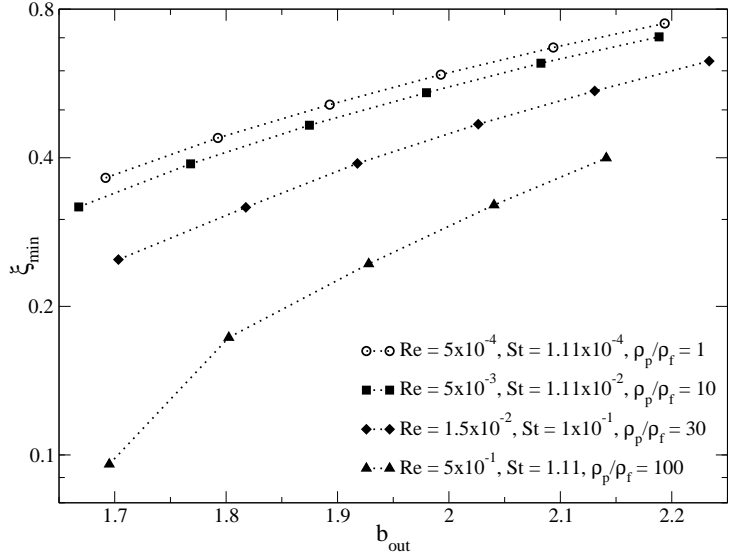}%{figureA10.pdf}%{../figures/ximinBout_bothReSt.pdf}
\caption{Effect of both significant fluid as well as particle inertia on the $\xi_{min}$--$b_{out}$ relationship, 
in the absence of pinching.}
\label{figbout:bothReStNoWallSupp}
\end{figure}

\begin{figure}
\includegraphics[width=\fw]{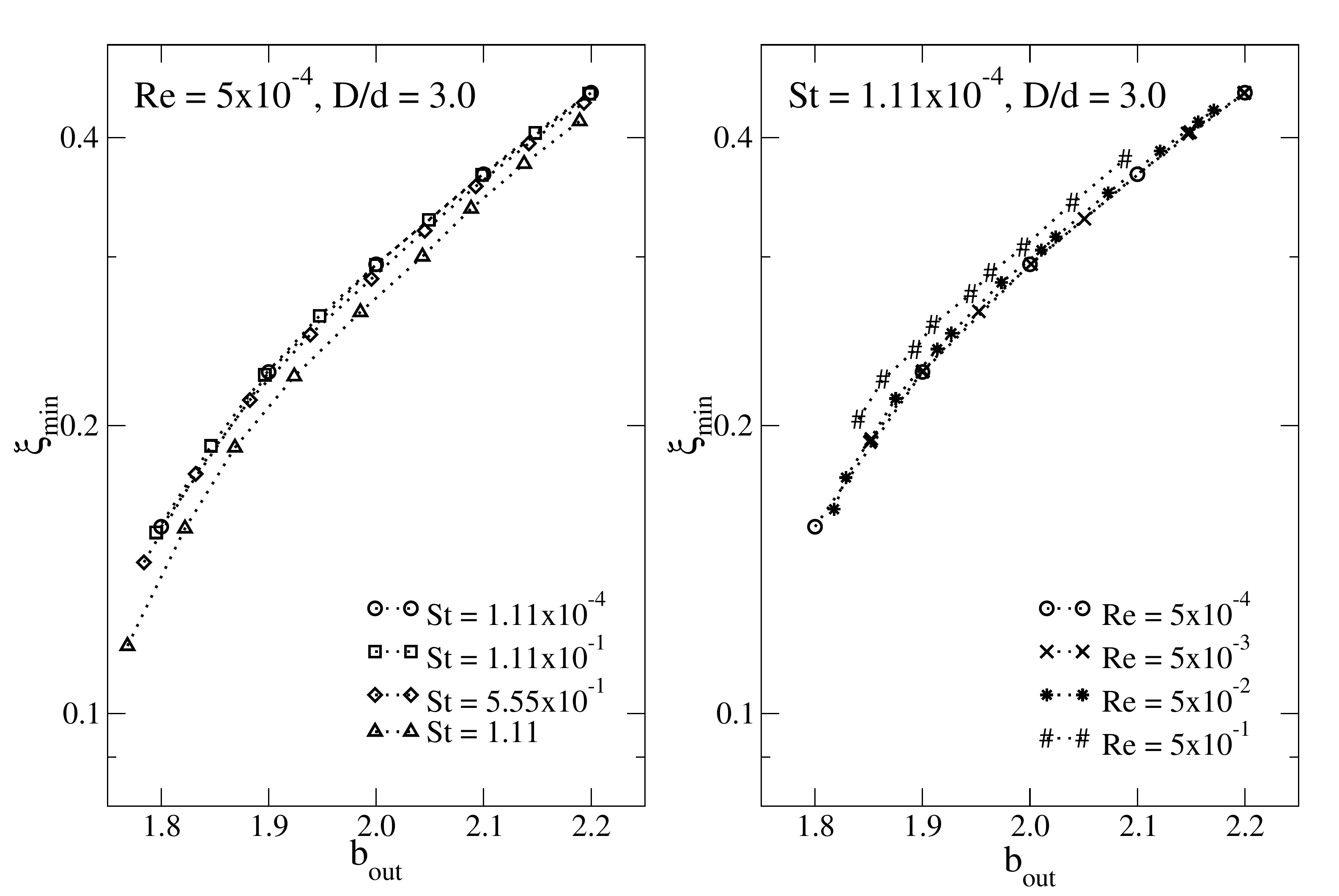}%{figureA11.pdf}%{../figures/ximinBout_Dbyd30wInertia.pdf}
\caption{The individual effect of particle and fluid inertia on the $\xi_{min}$--$b_{out}$ relationship in the presence of pinching. 
The aperture of the pinching gap is $3.0d$ (i.e., $d/D=0.333$). (Left) The effect of particle inertia, (right) the 
effect of fluid inertia.}
\label{figOut:bothInertiaNdWallat30}
\end{figure}

Consider figure \ref{fig:onlyWallNoInertia} in light of the hard-core model of the non-hydrodynamic repulsive interactions explained 
in the main text. For a fixed range of the interactions $\epsilon$, we have explained in the main text that the critical final-offset 
completely characterizes the lateral displacement of the particle downstream of the obstacle. Given that the case presented in figure 
\ref{fig:onlyWallNoInertia} corresponds to 
negligible inertia (Stokes regime), both critical initial- and final-offsets are equal. Thus, figure \ref{fig:onlyWallNoInertia} can 
also be interpreted as a plot of $\epsilon$ versus $b_c$, and clearly shows that the critical offset $b_c$ increases with 
a decreasing aperture of the pinching gap. Similar conclusions can be drawn concerning the critical initial-offset in 
absence of a pinching wall using figures \ref{fig:onlyInertiaNoWall} and \ref{fig:bothReStNoWall}, that is, the critical initial-offset 
increases for a fixed range of the repulsive interactions, with increasing magnitude of fluid and/or particle inertia.

\begin{figure}
\includegraphics[width=\fw]{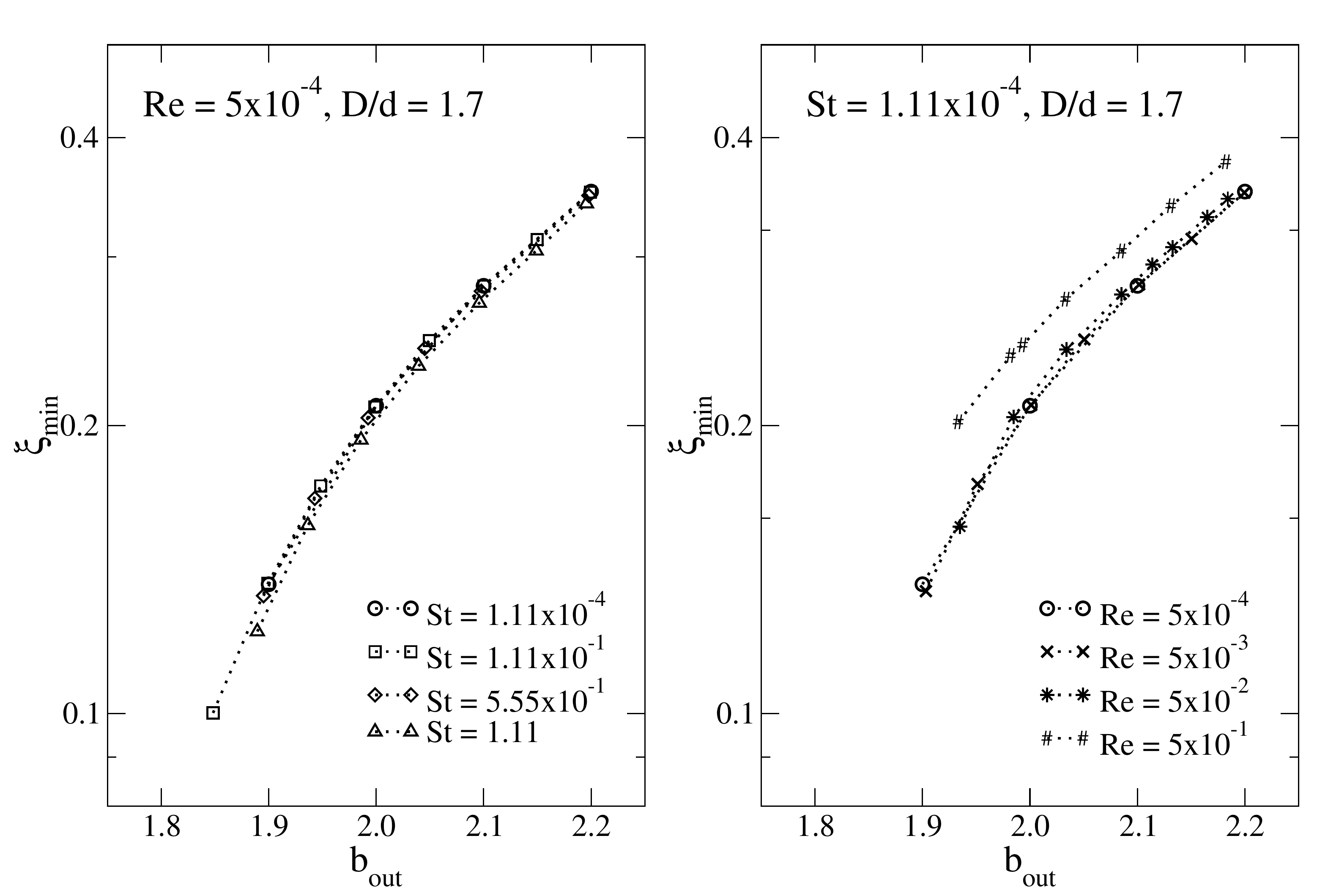}%{figureA12.pdf}%{../figures/ximinBout_Dbyd17wInertia.pdf}
\caption{The individual effect of particle and fluid inertia on the $\xi_{min}$--$b_{out}$ relationship in the presence of pinching. 
The aperture of the pinching gap is $1.7d$ (i.e., $d/D=0.588$). (Left) The effect of particle inertia, (right) the 
effect of fluid inertia.}
\label{figOut:bothInertiaNdWallat17}
\end{figure}

The asymmetry imparted to the particle trajectories by particle inertia, is readily seen in figure \ref{fig:boutBinReSttwoGraphs}. 
This figure supplements the information provided in figures $3$ and $5$ in the main text. We observe that, 
in the absence of a pinching wall, the final offset decreases with increasing particle inertia, while fluid inertia has a weak 
influence on it. 

\begin{figure}
\includegraphics[width=\fw]{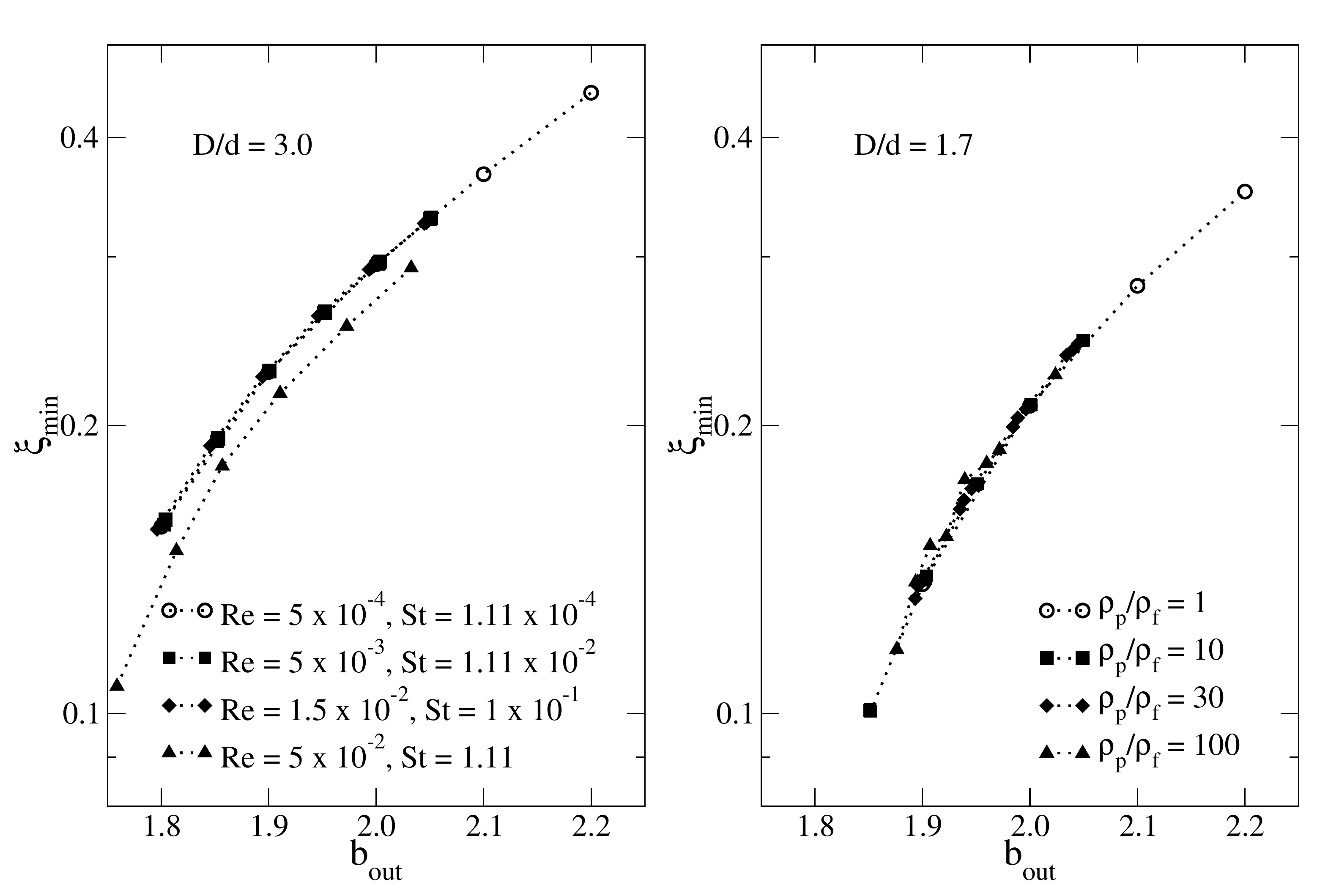}%{figureA13.pdf}%{../figures/ximinBout_bothDbyd_bothInertia.pdf}
\caption{The effect of inertia on the $\xi_{min}$--$b_{out}$ relationship, in the presence of pinching for two different apertures 
of the pinching gap. (Left) The aperture of the pinching gap is $3.0d$ (i.e., $d/D=0.333$), (right) the aperture is $1.7d$ (i.e., 
$d/D=0.588$).}
\label{figOut:bothInertiaNdWall}
\end{figure}

Figures \ref{figbout:onlyInertiaNoWallSupp} and \ref{figbout:bothReStNoWallSupp} present the effect of inertia on the minimum separation, 
as a function of the final offset, and complement figures \ref{fig:onlyInertiaNoWall} and \ref{fig:bothReStNoWall} respectively. 
These figures show that the critical final-offset increases with the magnitude of inertia for a constant range of the repulsive 
interactions, $\epsilon$.

The presence of both pinching as well as inertia poses the most complex case for analysis, especially, when non-hydrodynamic repulsive 
forces also play a role. Figures \ref{figOut:bothInertiaNdWallat30}, \ref{figOut:bothInertiaNdWallat17} and \ref{figOut:bothInertiaNdWall} 
are analogous to figures \ref{fig:bothInertiaNdWallat30}, \ref{fig:bothInertiaNdWallat17} and \ref{fig:bothInertiaNdWall}, respectively. 
Again, we see that the effect of pinching is to reduce the minimum separation and that the effect of inertia is small in all three figures.
However, in figures \ref{figOut:bothInertiaNdWallat30} and \ref{figOut:bothInertiaNdWallat17}, we observe a reversal of the usual trend 
when $Re\sim O(1)$, with high inertia leading to smaller values of the final offset. As discussed in the main text, this is caused by lift 
forces, and is therefore a case that should be considered separately.

\clearpage
\bibliographystyle{unsrtnat}

\begin{thebibliography}{30}
\providecommand{\natexlab}[1]{#1}
\providecommand{\url}[1]{\texttt{#1}}
\expandafter\ifx\csname urlstyle\endcsname\relax
  \providecommand{\doi}[1]{doi: #1}\else
  \providecommand{\doi}{doi: \begingroup \urlstyle{rm}\Url}\fi

\bibitem[Yamada et~al.(2004)Yamada, Nakashima, and Seki]{yamada2004}
M.~Yamada, M.~Nakashima, and M.~Seki.
\newblock Pinched flow fractionation: continuous size separation of particles
  utilizing a laminar flow profile in a pinched microchannel.
\newblock \emph{Anal. Chem.}, 76\penalty0 (18):\penalty0 5465--5471, 2004.

\bibitem[Xuan et~al.(2010)Xuan, Zhu, and Church]{xuan2010}
X.~Xuan, J.~Zhu, and C.~Church.
\newblock Particle focusing in microfluidic devices.
\newblock \emph{Microfluid. Nanofluid.}, 9:\penalty0 1--16, 2010.

\bibitem[Faivre et~al.(2006)Faivre, Abkarian, Bickraj, and Stone]{faivre2006}
M.~Faivre, M.~Abkarian, K.~Bickraj, and H.~A. Stone.
\newblock Geometrical focusing of cells in a microfluidic device: An approach
  to separate blood plasma.
\newblock \emph{Biorheology}, 43:\penalty0 147, 2006.

\bibitem[Wyss et~al.(2006)Wyss, Blair, Morris, Stone, and Weitz]{wyss2006}
H.~M. Wyss, D.~L. Blair, J.~F. Morris, H.~A. Stone, and D.~A. Weitz.
\newblock Mechanism for clogging of microchannels.
\newblock \emph{Phys. Rev. E}, 74:\penalty0 061402, 2006.

\bibitem[Mustin and Stoeber(2010)]{mustin2010}
B.~Mustin and B.~Stoeber.
\newblock Deposition of particles from polydisperse suspensions in microfluidic
  systems.
\newblock \emph{Microfluid. Nanofluid.}, 9:\penalty0 905--913, 2010.

\bibitem[Jain and Posner(2008)]{jain2008}
A.~Jain and J.~D. Posner.
\newblock Particle dispersion and separation resolution of pinched flow
  fractionation.
\newblock \emph{Anal. Chem.}, 80:\penalty0 1641--1648, 2008.

\bibitem[Mortensen(2007)]{mortensen2007}
N.~A. Mortensen.
\newblock Comment on “pinched flow fractionation: Continuous size separation
  of particles utilizing a laminar flow profile in a pinched microchannel”.
\newblock \emph{Anal. Chem.}, 79:\penalty0 9240--9241, 2007.

\bibitem[Shardt et~al.(2012)Shardt, Mitra, and Derksen]{shardt2012}
O.~Shardt, S.~K. Mitra, and J.~J. Derksen.
\newblock Lattice boltzmann simulations of pinched flow fractionation.
\newblock \emph{Chem. Eng. Sci.}, 75:\penalty0 106--119, 2012.

\bibitem[Balvin et~al.(2009)Balvin, Sohn, Iracki, Drazer, and
  Frechette]{balvin2009}
M.~Balvin, E.~Sohn, T.~Iracki, G.~Drazer, and J.~Frechette.
\newblock Directional locking and the role of irreversible interactions in
  deterministic hydrodynamics separations in microfluidic devices.
\newblock \emph{Phys. Rev. Lett.}, 103:\penalty0 078301, 2009.

\bibitem[Takagi et~al.(2005)Takagi, Yamada, Yasuda, and Seki]{takagi2005}
J.~Takagi, M.~Yamada, M.~Yasuda, and M.~Seki.
\newblock Continuous particle separation in a microchannel having
  asymmetrically arranged multiple branches.
\newblock \emph{Lab Chip}, 5:\penalty0 778--784, 2005.

\bibitem[Sai et~al.(2006)Sai, Yamada, Yasuda, and Seki]{sai2006}
Y.~Sai, M.~Yamada, M.~Yasuda, and M.~Seki.
\newblock Continuous separation of particles using a microfluidic device
  equipped with flow rate control valves.
\newblock \emph{J. Chromatogr. A}, 1127:\penalty0 214, 2006.

\bibitem[Maenaka et~al.(2008)Maenaka, Yamada, Yasuda, and Seki]{maenaka2008}
H.~Maenaka, M.~Yamada, M.~Yasuda, and M.~Seki.
\newblock Continuous and size-dependent sorting of emulsion droplets using
  hydrodynamics in pinched microchannels.
\newblock \emph{Langmuir}, 24:\penalty0 4405--4410, 2008.

\bibitem[Vig and Kristensen(2008)]{vig2008}
A.~L. Vig and A.~Kristensen.
\newblock Separation enhancement in pinched flow fractionation.
\newblock \emph{Appl. Phys. Lett.}, 93:\penalty0 203507, 2008.

\bibitem[Park et~al.(2009)Park, Song, and Jung]{park2009}
J.-S. Park, S.-H. Song, and H.-I. Jung.
\newblock Continuous focusing of microparticles using inertial lift force and
  vorticity via multi-orifice microfluidic channels.
\newblock \emph{Lab Chip}, 9:\penalty0 939--948, 2009.

\bibitem[Luo et~al.(2011)Luo, Sweeney, Risbud, Drazer, and Frechette]{luo2011}
M.~Luo, F.~Sweeney, S.~R. Risbud, G.~Drazer, and J.~Frechette.
\newblock Irreversibility and pinching in deterministic particle separation.
\newblock \emph{Appl. Phys. Lett.}, 99\penalty0 (6):\penalty0 064102, 2011.

%\bibitem[Sup()]{SuppInfo}
%{See supplementary information at {\em URL} for the simulation-data from the
%  perspective of the minimum separation $\xi_{min}$ as a function of the
%  initial offset $b_{in}$. This perspective brings more clarity to the
%  discussion of existence of critical offsets and their behavior in the
%  presence of inertia and a pinching wall.}
%
\bibitem[Ladd(1994{\natexlab{a}})]{ladd1994a}
A.~J.~C. Ladd.
\newblock Numerical simulations of particulate suspensions via a discretized
  boltzmann equation. part 1. theoretical foundation.
\newblock \emph{J . Fluid Mech.}, 271:\penalty0 285--309, 1994{\natexlab{a}}.

\bibitem[Ladd(1994{\natexlab{b}})]{ladd1994b}
A.~J.~C. Ladd.
\newblock Numerical simulations of particulate suspensions via a discretized
  boltzmann equation. part 2. numerical results.
\newblock \emph{J . Fluid Mech.}, 271:\penalty0 311--339, 1994{\natexlab{b}}.

\bibitem[Nguyen and Ladd(2002)]{nguyen2002}
N.-Q. Nguyen and A.~J.~C. Ladd.
\newblock Lubrication corrections for lattice-boltzmann simulations of particle
  suspensions.
\newblock \emph{Phys. Rev. E}, 66:\penalty0 046708, 2002.

\bibitem[Subramanian and Brady(2006)]{subramanian2006}
G.~Subramanian and J.~F.~Brady.
\newblock Trajectory analysis for non-Brownian inertial suspensions in simple shear flow.
\newblock \emph{J. Fluid Mech.}, 559:\penalty0 151--203, 2006.

\bibitem[Note1()]{Note1}
Note1.
\newblock It has been reported that using kinematic viscosity 
$\nu = 1/6$ (in lattice units) hastens the simulations as well as reduces numerical errors [Ladd, A. J. C. and Verberg, R. 
(2001), J Stat Phys, Vol.104, Nos. 5/6, pp. 1191--1251], providing a basis for our choice. Also, we note that the accuracy 
of the lattice-Boltzmann method based on the inherent compressibility, is reported to be $O(Ma^2)$ in the literature [e.g., 
Ladd, A. J. C. and Verberg, R. (2001), J Stat Phys, Vol.104, Nos. 5/6, pp. 1191-1251 and Hazi, G. (2003), Phys Rev E, Vol. 
67, article 056705], where $Ma$ is the Mach number associated with the system. For our system, the largest instantaneous 
particle velocities observed in the simulations are $O(10^{-2}~\text{lattice units})$. Given that the LB-code is D3Q19, 
velocity of sound is given by $c_s^2 = 1/3$ in lattice units. Thus, $Ma \sim O(10^{-2})$ and errors related to inherent 
compressibility of the method are $O(10^{-4})$.

\bibitem[Note2()]{Note2}
Note2.
\newblock We note that, in our simulations we specify the constant external
  acceleration acting on the particle, which is multiplied by the particle mass
  to compute the net force acting on it. Therefore, the expression for the
  Stokes settling velocity in equation \ref {eqn:defRe} contains just the
  particle density $\rho _p$, and not the difference between the particle and
  fluid densities, $\Delta \rho = \rho _p-\rho _f$.

\bibitem[Frechette and Drazer(2009)]{frechette2009}
J.~Frechette and G.~Drazer.
\newblock Directional locking and deterministic separation in periodic arrays.
\newblock \emph{J. Fluid Mech.}, 627:\penalty0 379--401, 2009.

\bibitem[Risbud and Drazer()]{risbud2013}
S.~R. Risbud and G.~Drazer.
\newblock Trajectory and distribution of non-brownian suspended particles
  moving past a fixed spherical or cylindrical obstacle.
\newblock \emph{J. Fluid Mech.}, 714:\penalty0 213--237, 2013.

\bibitem[Russel et~al.(1989)Russel, Saville, and Schowalter]{russel1989}
W.B. Russel, D.A. Saville, and W.R. Schowalter.
\newblock \emph{Colloidal dispersions}.
\newblock {Cambridge Monographs on Mechanics and Applied Mathematics}.
  Cambridge University Press, Cambridge, 1989.

\bibitem[Smart and {Leighton, Jr.}(1989)]{smart1989}
J.~R. Smart and D.~T. {Leighton, Jr.}
\newblock Measurement of the hydrodynamic surface roughness of noncolloidal
  spheres.
\newblock \emph{Phys. Fluids A}, 1\penalty0 (1):\penalty0 52--60, 1989.

\bibitem[Smart et~al.(1993)Smart, Beimfohr, and {Leighton, Jr.}]{smart1993}
J.~R. Smart, S.~Beimfohr, and D.~T. {Leighton, Jr.}
\newblock Measurement of the translational and rotational velocities of a
  noncolloidal sphere rolling down a smooth inclined plane at low reynolds
  number.
\newblock \emph{Phys. Fluids}, 5\penalty0 (1):\penalty0 13--24, 1993.

\bibitem[Russel(1980)]{russel1980}
W.~B. Russel.
\newblock Review of the role of colloidal forces in the rheology of
  suspensions.
\newblock \emph{J. Rheol.}, 24:\penalty0 287--317, 1980.

\bibitem[Rampall et~al.(1997)Rampall, Smart, and {Leighton, Jr.}]{rampall1997}
I.~Rampall, J.~R. Smart, and D.~T. {Leighton, Jr.}
\newblock The influence of surface roughness on the particle-pair distribution
  function of dilute suspensions of non-colloidal spheres in simple shear flow.
\newblock \emph{J. Fluid Mech.}, 339:\penalty0 1--24, 1997.

\bibitem[Brady and Morris(1997)]{brady1997}
J.~F. Brady and J.~F. Morris.
\newblock Microstructure of strongly sheared suspensions and its impact on
  rheology and diffusion.
\newblock \emph{J. Fluid Mech.}, 348:\penalty0 103--139, 1997.

\bibitem[da~Cunha and Hinch(1996)]{dacunha1996}
F.~R. da~Cunha and E.~J. Hinch.
\newblock Shear-induced dispersion in a dilute suspension of rough spheres.
\newblock \emph{J. Fluid Mech.}, 309:\penalty0 211--223, 1996.

\bibitem[Davis(1992)]{davis1992}
R.~H. Davis.
\newblock Effects of surface-roughness on a sphere sedimenting through a dilute
  suspension of neutrally buoyant spheres.
\newblock \emph{Phys. Fluids}, 4\penalty0 (12):\penalty0 2607--2619, 1992.

\end{thebibliography}

\end{document}